\documentclass[
aps,
twocolumn,
amsmath,
amssymb,
floatfix,
showpacs,
superscriptaddress,
nofootinbib,
longbibliography
]{revtex4-2}
\usepackage[T1]{fontenc}
\usepackage{mathtools}
\usepackage{braket}
\usepackage{bm}
\usepackage{MnSymbol}
\usepackage{graphicx}
\usepackage{tikz}
\usepackage{pgffor}
\usepackage{blkarray}
\usepackage{sidecap}

\usepackage[dvipsnames]{xcolor}
\usepackage{enumitem}
\usepackage[normalem]{ulem}
\usepackage{silence}

\usepackage[
colorlinks,
linkcolor=black,
citecolor=black,
urlcolor=black
]{hyperref}

\mathchardef\mhyphen="2D

\newcommand\bea{\begin{eqnarray}}
\newcommand\eea{\end{eqnarray}}
\newcommand\beq{\begin{equation}}
\newcommand\eeq{\end{equation}}

\definecolor{lime}{HTML}{A6CE39}

\DeclareRobustCommand{\orcidicon}{%
\hspace{-1.0mm}
\begin{tikzpicture}
  \draw[lime, fill=lime] (0,0)
  circle [radius=0.15]
  node[white] {{\fontfamily{qag}\selectfont \tiny \,ID}};
  \draw[white, fill=white] (-0.0525,0.095)
  circle [radius=0.007];
\end{tikzpicture}
\hspace{-3.0mm}
}

\foreach \x in {A,...,Z}{%
  \expandafter\xdef\csname orcid\x\endcsname{%
    \noexpand\href{https://orcid.org/\csname orcidauthor\x\endcsname}%
    {\noexpand\orcidicon}}%
}

\begin{document}
\title{Large nonlinear Hall effect in strained moiré structures hosting pseudospin-3/2 fermions}
\author{Srijata Lahiri$^\clubsuit$\orcidA{}}
\email[]{srijata.lahiri@iitg.ac.in} 
\affiliation{Department of Physics, Indian Institute of Technology Guwahati, Guwahati-781039, Assam, India}
\author{Gourab Paul$^\clubsuit$\orcidB{}}
\email[]{p.gourab@iitg.ac.in} 
\affiliation{Department of Physics, Indian Institute of Technology Guwahati, Guwahati-781039, Assam, India}
\author{Bilal Tanatar\orcidC{}}
\email[]{tanatar@fen.bilkent.edu.tr}
\affiliation{Department of Physics, Bilkent University, 06800 Bilkent, Ankara, Türkiye}
\author{Saurabh Basu}
\email[]{saurabh@iitg.ac.in}
\affiliation{Department of Physics, Indian Institute of Technology Guwahati, Guwahati-781039, Assam, India}

\def\thefootnote{$\clubsuit$}\footnotetext{These authors contributed equally to this work}

\begin{abstract} 
We investigate the linear and nonlinear Hall response of a moiré \emph{watermill lattice}, in which stacking and twisting generate a four-band manifold near the Fermi level with suppressed group velocities at discrete magic angles. Including an inversion symmetry breaking onsite mass breaks the interlayer symmetry, opening a gap in this manifold and driving the system into a non-trivial bulk topological phase. We map the resulting phase diagram as a function of the strength of the mass and twist angle $\theta$, revealing several sectors with high Chern numbers. We then introduce strain to break the residual $C_3$ symmetry of the lattice which activates a finite Berry curvature dipole and correspondingly, a nonlinear Hall response. The dipole reverses sign sharply across topological phase boundaries, producing butterfly like features when plotted against the relevant system parameters. Its magnitude substantially exceeds that reported for symmetry-broken transition metal dichalcogenides, consistent with the elevated Wilson-loop winding and enhanced quantum geometry associated with the lattice's pseudospin-$3/2$ character. 
We conclude by incorporating thermal effects on the Berry curvature dipole, asserting that it is an important tool for discerning topology at low temperatures.
\end{abstract}

\maketitle
\section{Introduction}
The recent studies of exotic quantum phenomena in long period moiré systems has emerged as a major research frontier in condensed matter physics~\cite{Cao2020,Cao2018,Tanaka2025,Ledwith2025,Mesple2025,Song2024,Yang2024,Devakul2021,Lei2025,Bao2025,Yang2024A,Lou2021,Sharma2021,Gandhi2026}. In twisted bilayers, the formation of nearly flat bands with suppressed Fermi velocity enhances electron-electron interactions, giving rise to a variety of correlated phases, including Mott insulating states~\cite{Po2018,Chen2021,Klug2020}, complex magnetic phases~\cite{Duran2022,Seo2019,Venderbos2018,Pan2020, Pan2020B,Keskiner2024}, and unconventional superconductivity~\cite{Balents2020,Isobe2018,Yankowitz2019,Oh2021,Bera2026}, among several others. Moiré physics in twisted bilayers of anisotropic lattices can also give rise to stripe-like and quasi one dimensional behavior, thereby leading to the emergence of density-wave and Luttinger-liquid phases~\cite{Wang2022,Wang2023,Kang2017,Santos2021,Wang2024,Hsu2023, Chang2025}. Besides the significant progress in the study of correlated physics, moiré systems have emerged as a fertile platform for exploring topological phenomena including anomalous quantum Hall (AQH)~\cite{Serlin2020,He2026,Tateishi2022} and fractional quantum Hall (FQH)~\cite{Xu2023,Reddy2023,Park2023} states.
\par Since the discovery of the quantum Hall effect (QHE) in 1979, the family of Hall phenomena has expanded considerably. A deeper understanding of the quantum origin of the transverse electron velocity has introduced the concepts of Berry curvature and topological Chern numbers into condensed matter physics~\cite{Xiao2010,Nagosa2010,Thouless1982}. Consequently, Hall responses have become powerful probes of the topology of the electronic band structures, as exemplified by the QHE and the AQH effect~\cite{Klitzing1980,Avron2003,Chang2013}. For AQH effect to exhibit a finite Hall signal, time reversal symmetry (TRS) of the system must be broken, either by applying an external magnetic field or through magnetic doping~\cite{Nagosa2010}. Remarkably, recent theoretical studies have predicted that a second-order Hall response, referred to as the nonlinear Hall effect (NLHE), can arise in time reversal symmetric materials when the inversion symmetry is broken~\cite{Du2018,Low2015,Zhang2018,Tokura2018,Zhang2022}. Subsequent experiments on bilayer and multilayer WTe$_2$ validated this prediction through the observation of a transverse nonlinear Hall-like current exhibiting a quadratic current-voltage characteristic~\cite{Ma2019,Kang2019}.
The microscopic origin of the NLHE lies in the non-vanishing dipole moment of the Berry curvature distribution in momentum space, known as the Berry curvature dipole (BCD)~\cite{Sodemann2015}. From an experimental perspective, the NLHE provides an effective probe of topological phase transitions, particularly in time reversal symmetric systems where the conventional Berry curvature induced Hall response vanishes. In particular, the nonlinear Hall (NLH) response exhibits a characteristic sign reversal across the phase boundaries, providing an efficient means to identify band inversion associated with gap closing in topological systems~\cite{Zhang2022,Chakraborty2022,Hu2022}. 
\par A wide range of materials has been theoretically predicted or experimentally demonstrated, which exhibit strong NLHE, including strained WSe$_2$~\cite{Qin2021}, corrugated graphene~\cite{Ho2021}, Weyl semimetals~\cite{Kumar2021,Zhang2018B,Singh2020,Tiwari2021}, MoTe$_2$~\cite{Ma2022}, and the giant Rashba material bismuth tellurium iodide (BiTeI) under applied pressure~\cite{Facio2018}.
Twisted bilayer systems also provide a promising platform for engineering NLHE that extend beyond the intrinsic properties of their constituent layers. Experimentally, NLH responses have been observed in several moiré superlattices, such as twisted bilayer graphene (TBG)~\cite{Duan2022,Huang2023}, twisted double bilayer graphene (TDBLG)~\cite{Sinha2022,Zhong2024}, twisted bilayer MoS$_2$~\cite{Wu2023}, and twisted bilayer WSe$_2$~\cite{Huang2023A,Cao2025}.
From a theoretical perspective, the NLH response in a material can be investigated by calculating the BCD from its electronic band structure. Such analyses have predicted the emergence of the NLHE in a variety of moiré systems, including strained TBG~\cite{Zhang2022}, twisted bilayer dice lattice (TBDL)~\cite{paul2026arXiv}, TDBLG~\cite{Chakraborty2022}, twisted bilayer WTe$_2$~\cite{He2021}, and twisted bilayer WSe$_2$~\cite{Hu2022}. It should be noted here that moiré lattice systems, particularly twisted bilayer graphene under strain, exhibits noticeably higher magnitudes of BCD and hence NLHE signatures, owing to enhanced quantum geometric effects within their flat band subspace. 
\par 
In this work, we deviate from conventional moiré graphene, a pseudospin-1/2 system, to study the emergence and modification of topological phases, and strain-driven nonlinear Hall (NLH) response, in a twisted bilayer moiré system hosting higher-pseudospin fermions.
Our central motivation is to test whether a higher-pseudospin lattice structure significantly affects the emergent topological and nonlinear-response features.
In this regard, we explore both the unstrained as well as the strained counterparts of a \textit{twisted bilayer} Watermill (WM) lattice which is known to host low energy states of pseudospin-$3/2$~\cite{Hung2025}. It should be mentioned that other systems, including the Dice lattice~\cite{Zhou2024,Ma2024,Gandhi2026,paul2026arXiv,Paul2026}, that digress from hosting low energy Dirac fermions, have been thoroughly explored in the literature. Importantly however, the monolayer Dice lattice being a three-band model~\cite{Illes2015,Illes2017,Islam2024,Islam2025,Mondal2023,Bhattacharyya2024}, leads to the formation of a highly degenerate flat band at all magic angles, thereby making the topology of the system complicated. Monolayer WM is however a four band model with an augmented hexagonal structure, which leads to the formation of a four band subspace near the Fermi level under stacking and twisting, enabling the effect of higher pseudospin on the topology as well NLHE, to be studied in a much less ambiguous setting. Furthermore, owing to the elevated Wilson loop windings and quantum geometric effects of the WM lattice \cite{Hung2025}, it is expected to be a promising platform for enhanced NLHE signatures. As a final objective, we also demonstrate that the BCD acts as a tool to probe topological phase boundaries in the strained counterpart of the twisted bilayer Watermill lattice, with its sign reversals tracking the associated band inversions.
\par With this motivation, the remainder of this work is organized as follows. We first introduce the unstrained \textit{twisted bilayer} WM lattice with the corresponding monolayer Hamiltonian discusssed in Appendix-\ref{Appendix-A}. We then systematically incorporate strain into the Hamiltonian, which breaks the relevant crystal symmetry and enables a finite NLH response. The resulting violation of $\mathcal{C}_{3z}$ symmetry is confirmed by examining the band structure along the high-symmetry path of the moiré Brillouin zone (MBZ). Again, elaborate discussion on the effect of strain in the intra and interlayer Hamiltonians have been given in the Appendix-\ref{Appendix-B}. We then analyze the Chern number phase diagrams of the two valence bands closest to the Fermi energy, first in the $\Delta$-$\theta$ plane under weak uniaxial strain ($\epsilon_p$), where $\Delta$ corresponds to the strength of the onsite mass, and subsequently in the $\Delta$-$\epsilon_p$ plane to examine the strain driven evolution of the topological phases. The simultaneous breaking of inversion and $\mathcal{C}_{3z}$ symmetries gives rise to a finite NLHE, which exhibits characteristic sign change across the topological phase boundaries. We subsequently demonstrate how the NLH response can serve as a probe for identifying and distinguishing the different topological phases obtained from the Chern number phase diagrams. We also examine the evolution of the BCD components with increasing strain strength, highlighting the essential role of $\mathcal{C}_{3z}$ symmetry violation in generating a finite second-order Hall response. Finally, we study the temperature dependence of the NLH response in both the chiral limit and the broken chiral-symmetry regime.
\section{Unstrained Hamiltonian}
\begin{figure}[h]
    \centering
    \includegraphics[width=0.6\linewidth]{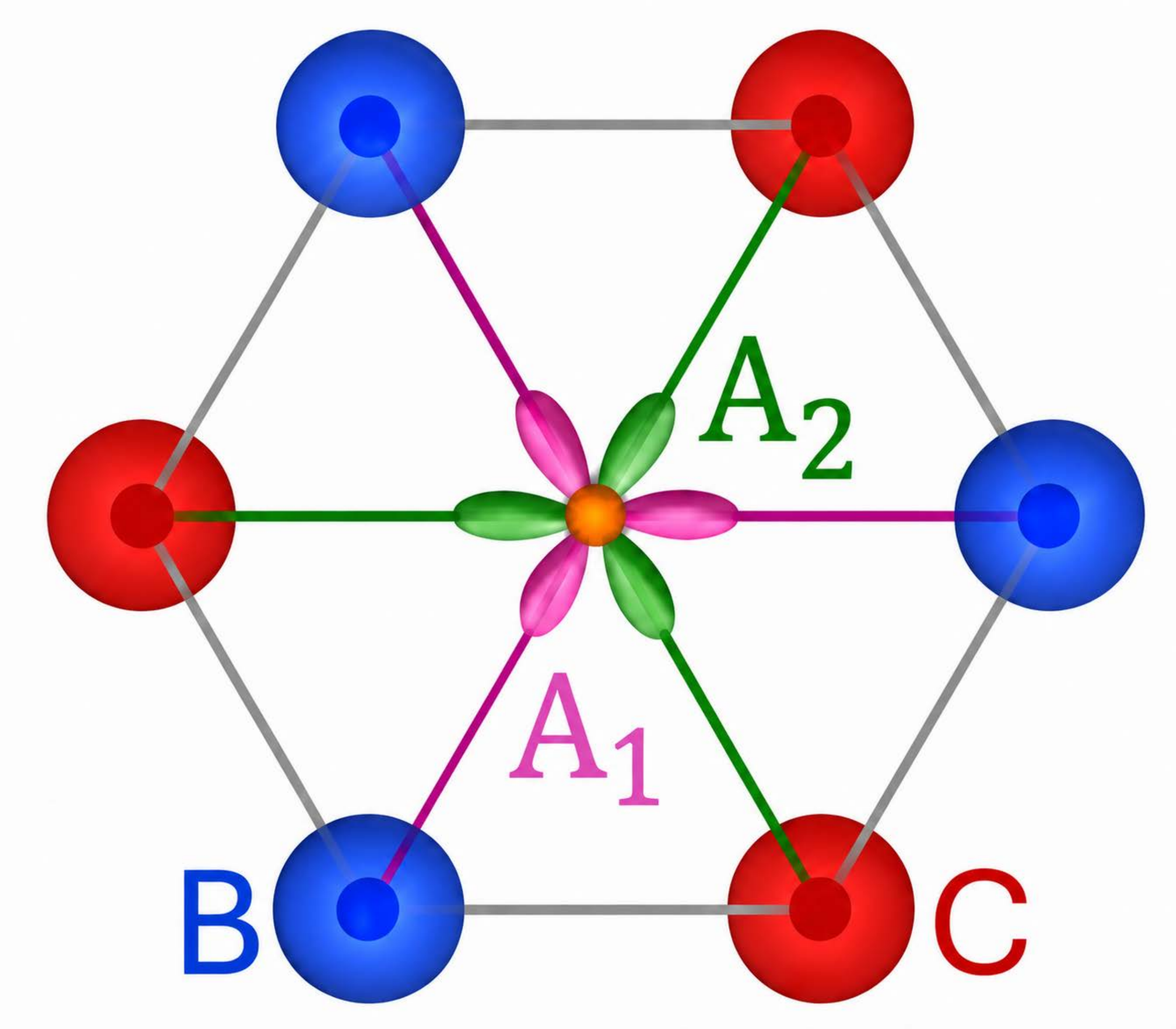}
    \caption{Schematic illustration of the WM lattice. The NN hopping between the $B$ and $C$ orbitals is represented by gray bonds, whereas the NN hopping between the $A_1$--$B$ and $A_2$--$C$ orbitals is highlighted by pink and green bonds, respectively.}
    \label{fig:Figure1}
\end{figure}

We begin by considering a single layer WM structure, which is inspired by the dice lattice and consists of a honeycomb framework augmented by a central atom $A$ hosting two distinct orbitals, denoted by $A_1$ and $A_2$. In contrast, each of the $B$ and $C$ sublattice sites forming the underlying honeycomb lattice accommodates only a single orbital. We consider only the nearest-neighbor (NN) hopping processes, where the electronic coupling between the central $A$ site and its neighboring $B$ ($C$) sites is mediated exclusively through the $A_1$ ($A_2$) orbital. The corresponding lattice geometry and hopping pathways are illustrated in Fig.~\ref{fig:Figure1}. It is important to emphasize that while the $B$ and the $C$ site are identical to each other, the central $A$ site is distinct owing to the presence of two active orbitals. The nearest-neighbor hopping amplitude between the $B$ and $C$ sites is denoted by $t$. Due to the presence of sixfold rotational symmetry ($\mathcal{C}_{6z}$), the hopping amplitudes between the $A_1$ and $B$ sites and between the $A_2$ and $C$ sites are considered to be symmetry equivalent. These hopping amplitudes are considered to be different in magnitude from the $B$-$C$ hopping and are taken to be equal to $t^{\prime}=\frac{\sqrt{3}}{2}t$, throughout our work. Extensive calculations pertaining to the monolayer WM lattice and its low-energy description have been presented in Appendix-\ref{Appendix-A}.
\par Having established the monolayer description, we now shift our attention to the composite bilayer of the WM lattice, where the top layer (layer-1) is rotated by an angle $+\theta/2$ and the bottom layer (layer-2) is rotated by an angle $-\theta/2$, resulting in a relative twist angle $\theta$ between the two layers. The corresponding continuum model for the \textit{twisted bilayer} WM lattice can be constructed directly following the 
Bistritzer-MacDonald formalism~\cite{Bistritzer2011} as,
\begin{equation}
    H^{\zeta}(\theta) = 
\begin{pmatrix}
H_{t}^\zeta\left( \frac{\theta}{2} \right) & T_{\zeta} \\
T_{\zeta}^\dagger & H_{b}^\zeta\left( -\frac{\theta}{2} \right)
\end{pmatrix}.
\label{Twisted_Bilayer_Ham}
\end{equation}
Here, $H^\zeta_{t/b}$ denotes the low-energy intralayer Hamiltonian for the top/bottom layer corresponding to the valley $\zeta$, which can be expressed as,
\begin{eqnarray}
 H_{t/b}^\zeta ({\theta/2})= v_F \mathcal{R}_{\mp \theta/2} \boldsymbol{q}.(\zeta S_1,S_2),   
\end{eqnarray}
where $v_F = 5960$~meV$\cdot$\AA,
and $\mathcal{R}_{\theta}$ denotes the rotation operator. The momentum $\mathbf{q} = \mathbf{k} - \mathbf{K}_{\zeta}$ is defined relative to the original Brillouin zone (BZ) corner $\mathbf{K}_{\zeta}$ 
of the rotated monolayer WM lattice and $S_i$ ($i = 1, 2, 3$) denotes the matrix representation of the spin-3/2 operators in the $S_3$ eigenbasis. Furthermore, the interlayer Hamiltonian $T_{\zeta}$, with the NN hopping from sublattice site $X$ ($X \in A_1, B, C, A_2$) in layer-1 (top layer) to the nearest sublattice site $\tilde{X}$ ($\tilde{X} \in \tilde{A}_1, \tilde{B}, \tilde{C}$, $\tilde{A}_2$) in layer-2 (bottom layer) is given by~\cite{Paul2026, Koshino2015},
\begin{eqnarray}
    T_\zeta(\mathbf{q}, \tilde{\mathbf{q}}) = T_{\zeta\mathbf{q}_b } \delta_{\mathbf{q}-\tilde{\mathbf{q}}- \zeta \mathbf{q}_b} + T_{\zeta \mathbf{q}_{tr}} \delta_{\mathbf{q}-\tilde{\mathbf{q}}- \zeta \mathbf{q}_{tr}} + T_{ \zeta\mathbf{q}_{tl}} \delta_{\mathbf{q}-\tilde{\mathbf{q}}- \zeta \mathbf{q}_{tl}},\label{Interlayer_ham} \nonumber \\\label{top_bottom_Ham}
\end{eqnarray}
where the momentum transfer vectors connecting the nearest Dirac points of the two layers in the MBZ are given by, $\textbf{q}_b$, $\textbf{q}_{tr}$, and $\textbf{q}_{tl}$. The matrices $T_{\zeta \mathbf{q}_b}$, $T_{\zeta \mathbf{q}_{tr}}$, and $T_{\zeta \mathbf{q}_{tl}}$, which describe the interlayer coupling associated with each momentum transfer in the $A$--$A$ stacking configuration, take the following forms, 
\begin{eqnarray}
T_{\zeta \mathbf{q}_b} &=&  \begin{pmatrix}
u_1 & u_2' & u_3 & u_4\\
u_2' & u_1 & u_2 & u_3\\
u_3 & u_2 & u_1 & u_2'\\
u_4 & u_3 & u_2' & u_1
\end{pmatrix}, \nonumber \\
T_{\zeta \mathbf{q}_{tr}} &=& \begin{pmatrix}
u_1 & u_2'e^{-i \zeta \phi} & u_3e^{i \zeta \phi} & u_4\\
u_2'e^{i \zeta \phi} & u_1 & u_2e^{-i\zeta \phi} & u_3e^{i\zeta \phi}\\
u_3e^{-i\zeta\phi} & u_2e^{i\zeta\phi} & u_1 & u_2'e^{-i\zeta\phi}\\
u_4 & u_3e^{-i\zeta\phi} & u_2'e^{i\zeta\phi} & u_1
\end{pmatrix}, \nonumber \\
T_{\zeta \mathbf{q}_{tl}} &=& \begin{pmatrix}
u_1 & u_2'e^{i \zeta \phi} & u_3e^{-i \zeta \phi} & u_4\\
u_2'e^{-i \zeta \phi} & u_1 & u_2e^{i\zeta \phi} & u_3e^{-i\zeta \phi}\\
u_3e^{i\zeta\phi} & u_2e^{-i\zeta\phi} & u_1 & u_2'e^{i\zeta\phi}\\
u_4 & u_3e^{i\zeta\phi} & u_2'e^{-i\zeta\phi} & u_1
\end{pmatrix}.  \label{interlayer_Ham}
\end{eqnarray}
\begin{figure}[h]
    \centering
    \includegraphics[width=0.5\linewidth]{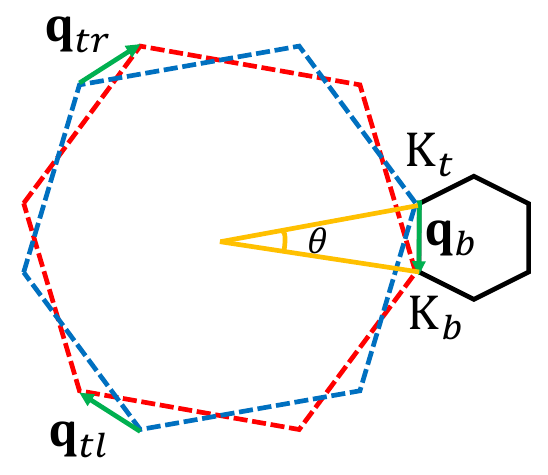}
    \caption{Momentum space structure of \textit{twisted bilayer} WM lattice. The dashed blue and red hexagons correspond to the Brillouin zones of the top and bottom WM lattice. The small black hexagon denotes the MBZ generated by the relative rotation $\theta$ between the two layers. The vectors $\mathbf{q}_{b}$, $\mathbf{q}_{tr}$, and $\mathbf{q}_{tl}$ represent the three momentum transfer vectors that couple the electronic states between the two layers.
}
    \label{fig:Figure12}
\end{figure}
\begin{figure}[h]
    \centering
    \includegraphics[width=1.0\linewidth]{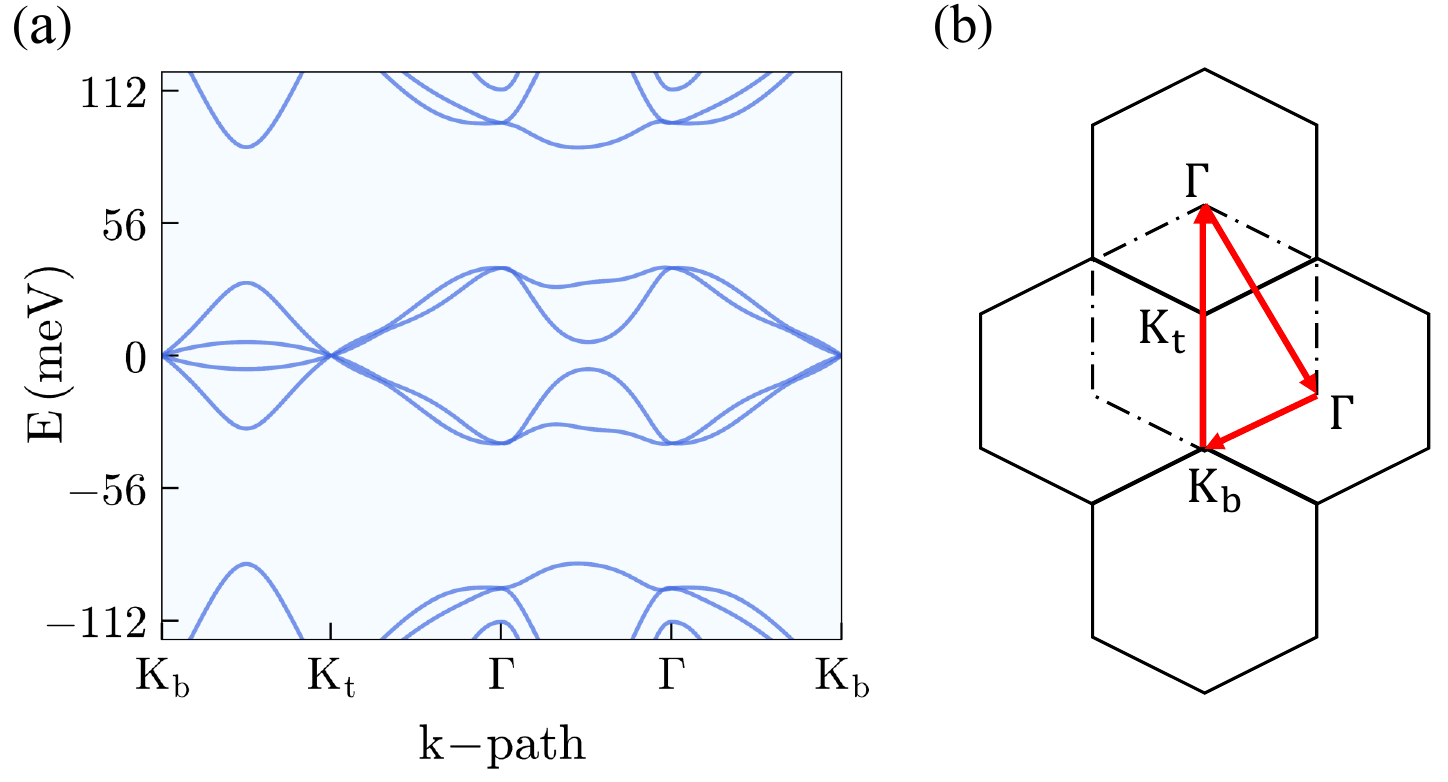}
    \caption{(a) Band structure of the \textit{twisted bilayer} WM lattice around $K_+$ valley at a twist angle of $\theta=1.40^\circ$ in the chiral limit of the system. (b) High-symmetry path in the MBZ.}
    \label{fig:Figure9}
\end{figure}
Here, $\phi=2\pi/3$. The parameters $u_{1,2,3,4}$ together with $u_2^{\prime}$ characterize the different interlayer hopping processes. Specifically, $u_1$ describes the intrasublattice coupling between the $A_1$-$\tilde{A}_1$, $B$-$\tilde{B}$, $C$-$\tilde{C}$, and $A_2$-$\tilde{A}_2$ orbitals, while $u_2$ accounts for the intersublattice tunneling between the $B$ and $\tilde{C}$ orbitals and their symmetry related counterparts. In addition, $u_3$ captures the interlayer hopping between the $A_1$ and $\tilde{C}$ orbitals as well as between the $A_2$ and $\tilde{B}$ orbitals, together with the corresponding reverse processes. Finally, $u_2^{\prime}$ governs the interlayer coupling between the $A_1$-$\tilde{B}$ and $A_2$-$\tilde{C}$ orbitals, whereas $u_4$ describes the tunneling between the $A_1$ and $\tilde{A}_2$ orbitals, and vice versa. The precise values of the interlayer hopping parameters. $u_{1,2,3,4}$ and $u_2^{\prime}$, are determined by the orbital character of the electrons on each sublattice. As a representative example of the \textit{twisted bilayer} WM lattice with sublattice symmetry, we consider the $A_1$ and $A_2$ orbitals are of \textit{sp$_2$} type, while the $B$ and $C$ orbitals possess $s$ character, thereby preserving the $\mathcal{C}_{3z}$ symmetry. This orbital arrangement establishes the following hierarchy of the interlayer hopping amplitudes: $u_1 \sim u_2 \sim u_2^{\prime} \sim u_3 \gg u_4$~\cite{Hung2025}.
\par The chiral limit of the \textit{twisted bilayer} WM lattice can be realized by adopting a unified hopping scheme for both the intralayer and interlayer couplings, with $u_1=u_3=u_4=0$, $u_2=110.7$~meV, and $u_2^{\prime}=\frac{\sqrt{3}}{2}u_2$. With these choices, the interlayer tunneling between the $A_1$-$B$, $A_2$-$C$, and $B$-$C$ orbital pairs follows the same hopping pattern as their corresponding intralayer processes, thereby preserving a unified hopping framework across the two layers. Now we examine the electronic band structure of the \textit{twisted bilayer} WM lattice around the $K_+$ valley in the chiral limit along the high-symmetry path, as shown in Fig.~\ref{fig:Figure9}(a). The energy spectrum exhibits four isolated, degenerate, nearly flat bands located close to the Fermi energy.
\section{Making moiré WM lattice topological} The origin and the topological nature of these nearly flat isolated bands become more evident in the chiral limit.
As seen in Fig.~\ref{fig:Figure9}(a), the subspace near the Fermi level becomes four-fold degenerate at the high symmetry points in the Brillouin zone which makes any topological characterization ill-defined~\cite{Pantaleon2021}.
With the objective of lifting the degeneracy of these four isolated bands, we now introduce an inversion symmetry breaking mass term on the  $A_1$ and $A_2$ orbitals, while the $B$ and $C$ orbitals remain unaffected. We have included the aforementioned effect in our calculation through the gap parameters as, $\Delta_{A_1}^t = \frac{3}{2} \Delta$, $\Delta_{A_2}^t = \frac{1}{2} \Delta$, $\Delta_{\tilde{A}_1}^b = -\frac{1}{2} \Delta$, and $\Delta_{\tilde{A}_2}^b = -\frac{3}{2} \Delta$, which emulates a constant potential gradient. The explicit form of the intralayer Hamiltonian in the presence of this mass configuration is provided in Appendix-\ref{Appendix-B}.
\begin{figure}[h]
    \centering
    \includegraphics[width=0.68\linewidth]{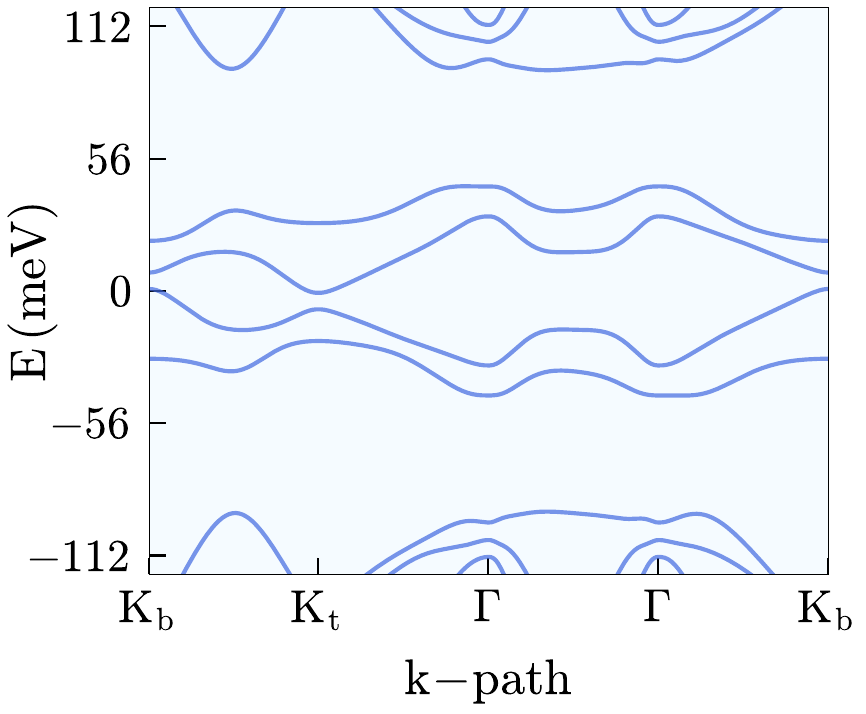}
    \caption{Band structure of the \textit{twisted bilayer} WM lattice around the $K_+$ valley in the presence of an onsite mass ($\Delta = 38$~meV), calculated at a twist angle of $\theta = 1.40^\circ$ in the chiral limit of the system along the high-symmetry path shown in Fig.~\ref{fig:Figure9}(b).}
    \label{fig:Figure10}
\end{figure}
\par
\begin{figure}[]
    \centering
    \includegraphics[width=1.0\linewidth]{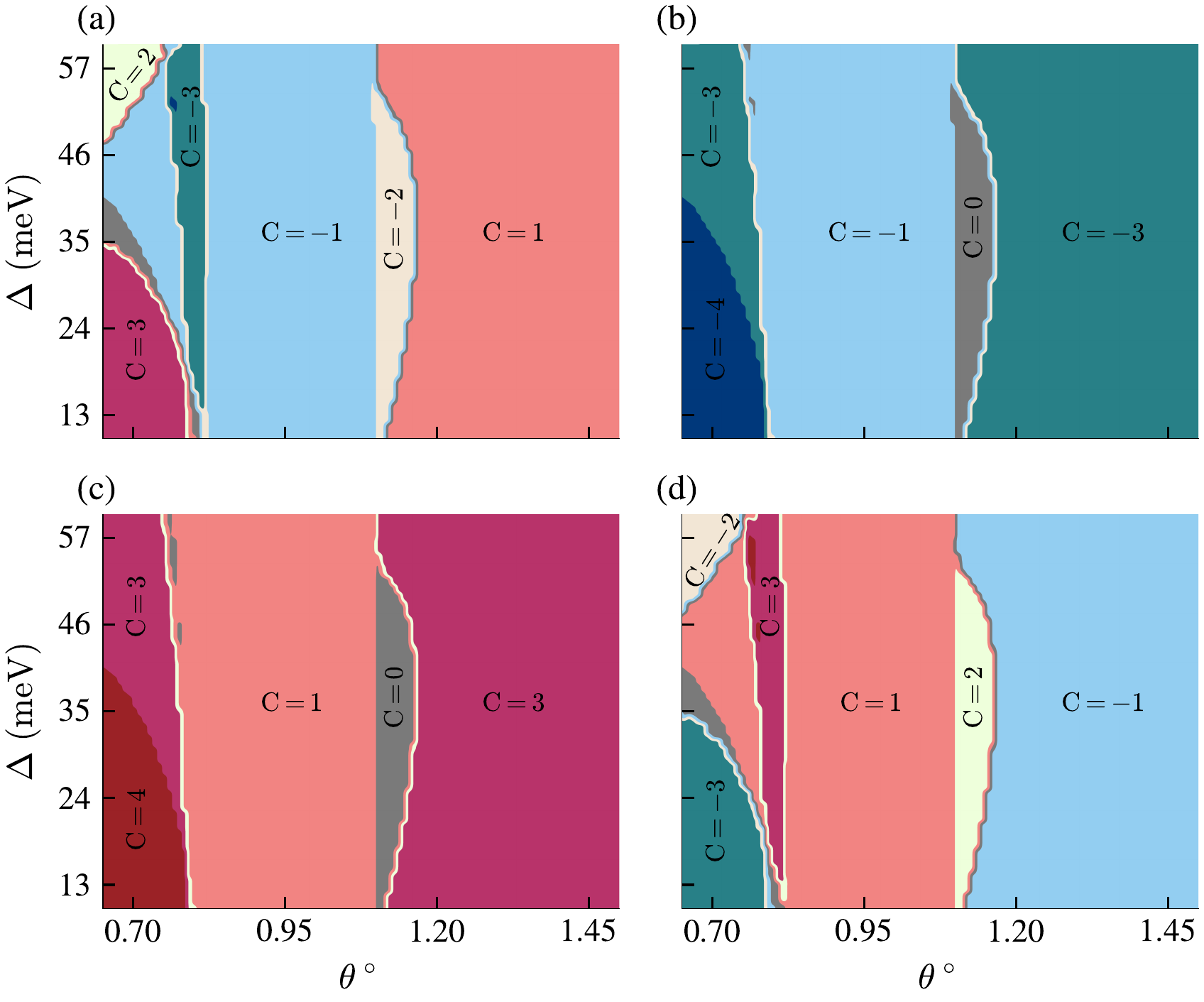}
    \caption{Unstrained Chern phase diagrams corresponding to the four isolated nondegenerate subbands close to the charge neutrality. Panels (a)–(d) correspond to band indices $n=322$, $323$, $324$, and $325$, respectively, and illustrate the evolution of the Chern number as a function of onsite parameter $\Delta$ and twist angle $\theta$.}
    \label{fig:Figure8}
\end{figure}

 With this configuration, the inversion ($\mathcal{C}_{2z}$) symmetry of the system is broken, which give rise to four isolated, nondegenerate bands in the vicinity of the Fermi energy, as shown in Fig.~\ref{fig:Figure10}. We now commence our analysis by examining the topological phases of the unstrained \textit{twisted bilayer} WM lattice around the $K_+$ valley by mapping the Chern number ($C$) in the $\Delta$-$\theta$ parameter space. In Figs.~\ref{fig:Figure8}[(a)–(d)], we present the Chern number phase diagrams for the four isolated nondegenerate middle subbands closest to the charge neutrality point (band indices $n=322$, $323$, $324$, and $325$) in the chiral limit of the system \cite{Note}. The phase diagrams reveal several distinct topological regions characterized by Chern numbers $C=0$, $\pm1$, $\pm2$, $\pm3$, and $\pm4$, separated by distinct transition lines. These topological phase transitions are associated with the closure and reopening of the band gap at specific $\mathbf{q}$ points in the MBZ. Owing to the valley-dependent nature of the MBZ, the system effectively lacks TRS within an individual valley, allowing nontrivial Berry curvature to develop once inversion symmetry is broken by the onsite parameter $\Delta$. As a result, $\Delta$ serves as an effective tuning parameter that drives a metal--insulator transition without changing the carrier doping, while concurrently triggering a valley-Chern topological phase transition.
\par However, it is important to emphasize here that the breaking of TRS is a valley-resolved property and does not extend to the complete system. When the contributions from both the $K_+$ and $K_-$ valleys are considered, the full Hamiltonian preserves TRS, causing the opposite Berry curvature contributions to cancel exactly and resulting in a vanishing total Chern number. As a result, the bulk band geometry and the associated topological phase transitions are challenging to detect experimentally, since the vanishing total Chern number precludes the existence of net circulating edge states. Nevertheless, when the crystal symmetry is reduced through the breaking of $\mathcal{C}_{2z}$ and $\mathcal{C}_{3z}$ symmetries, time reversal symmetric systems exhibit emergence of NLHE owing to non-vanishing BCD. Thus, BCD serves as an effective tool for detecting topological phase transitions in time reversal symmetric systems.
\section{Strained Hamiltonian}
As discussed earlier, the emergence of a finite BCD in a time reversal symmetric system requires the reduction of the crystal symmetry, particularly the breaking of $\mathcal{C}_{2z}$ and $\mathcal{C}_{3z}$ symmetries. In the \textit{twisted bilayer} WM lattice, we break the $\mathcal{C}_{3z}$ symmetry by introducing a uniaxial strain, which shifts the location of Dirac cones in momentum space without gapping them out. Strain is commonly encountered in two-dimensional (2D) moiré systems as a result of substrate induced effects. In twisted moiré structures, where two layers are stacked and encapsulated between substrates, strain can in general be present in both the layers~\cite{Huang2022, Pantaleon2022, Ouyang2025, Hou2025}. However, for the simplicity in our calculation, we assume that strain acts only on the bottom layer of the system~\cite{Zhang2022, paul2026arXiv}. Now to incorporate the effect of strain in the bottom layer Hamiltonian, we introduce a linear strain tensor $\mathbf{\mathcal{E}}$, which modifies the phase space coordinates ($\mathbf{r}$, $\mathbf{q}$) as~\cite{Bi2019},
\begin{eqnarray}
    \mathbf{r}^\prime &=& (\mathbb{I} + \mathbf{\mathcal{E}})\,\mathbf{r} \nonumber \\
    \mathbf{q}^\prime &=& (\mathbb{I} + \mathbf{\mathcal{E}^T})^{-1}\,\mathbf{q} \approx (\mathbb{I} - \mathbf{\mathcal{E}^T})\,\mathbf{q}.
\end{eqnarray}
Without any loss of generality, we consider a uniaxial strain of magnitude $\epsilon_p$ applied along a direction oriented at an angle $\beta$ with respect to the zigzag axis ($\beta=0$). The corresponding strain tensor can be written as~\cite{Bi2019, Pereira2009},
\begin{eqnarray}
\mathbf{\mathcal{E}}
&=&
R_{\beta}
\begin{pmatrix}
\epsilon_p & 0 \\
0 & -\nu \epsilon_p
\end{pmatrix}
R_{\beta}^{-1}
\nonumber \\
&=&
\epsilon_p
\begin{pmatrix}
\cos^2\beta-\nu\sin^2\beta &
(1+\nu)\cos\beta\sin\beta \\
(1+\nu)\cos\beta\sin\beta &
\sin^2\beta-\nu\cos^2\beta
\end{pmatrix}.
\end{eqnarray}
Here, $\nu$ denotes the Poisson ratio. For the WM lattice, we use $\nu=0.165$, which is consistent with the value reported for graphene~\cite{Pereira2009}. The effect of strain on the bottom layer Hamiltonian introduces an effective gauge field~\cite{Sun2022, Bi2019}, which can be expressed as,
\begin{equation}
\mathbf{A} = \frac{\gamma}{d}
\left( \mathcal{E}_{xx} - \mathcal{E}_{yy}\,, -2\mathcal{E}_{xy} \right),
\end{equation}
Here, $\gamma=1.57$ denotes the Grüneisen parameter~\cite{Bi2019}, which we also assume to be same as in graphene. In the presence of this effective gauge field, the Dirac points of the bottom layer are shifted to~\cite{Bi2019},
\begin{equation}
\mathbf{\mathcal{B}}_{\zeta} = \left(\mathbb{I} - \mathbf{\mathcal{E}}^{T}\right)\,\mathbf{K}_{b,\zeta} - \zeta\,\mathbf{A}. 
\end{equation}
Now including the potential gradient $V_b$ (captures the effect of onsite potential in the bottom layer), the bottom layer Hamiltonian takes the form,
\begin{eqnarray}
 H_{b}^{\prime \zeta} =  v_F \,{R}_{\theta/2} \left[ (\mathbb{I} + \mathbf{\mathcal{E}}^T)\,\mathbf{q}^{\prime} + \zeta \mathbf{A} \right].(\zeta S_1, S_2) + V_b.
\end{eqnarray}
It is important to note that strain also influences the interlayer couplings by modifying the momentum transfers associated with the interlayer tunneling processes. A comprehensive discussion of the strain modified intralayer and interlayer Hamiltonians is provided in Appendix-\ref{Appendix-B}.
\begin{figure}[h]
    \centering
    \includegraphics[width=1.0\linewidth]{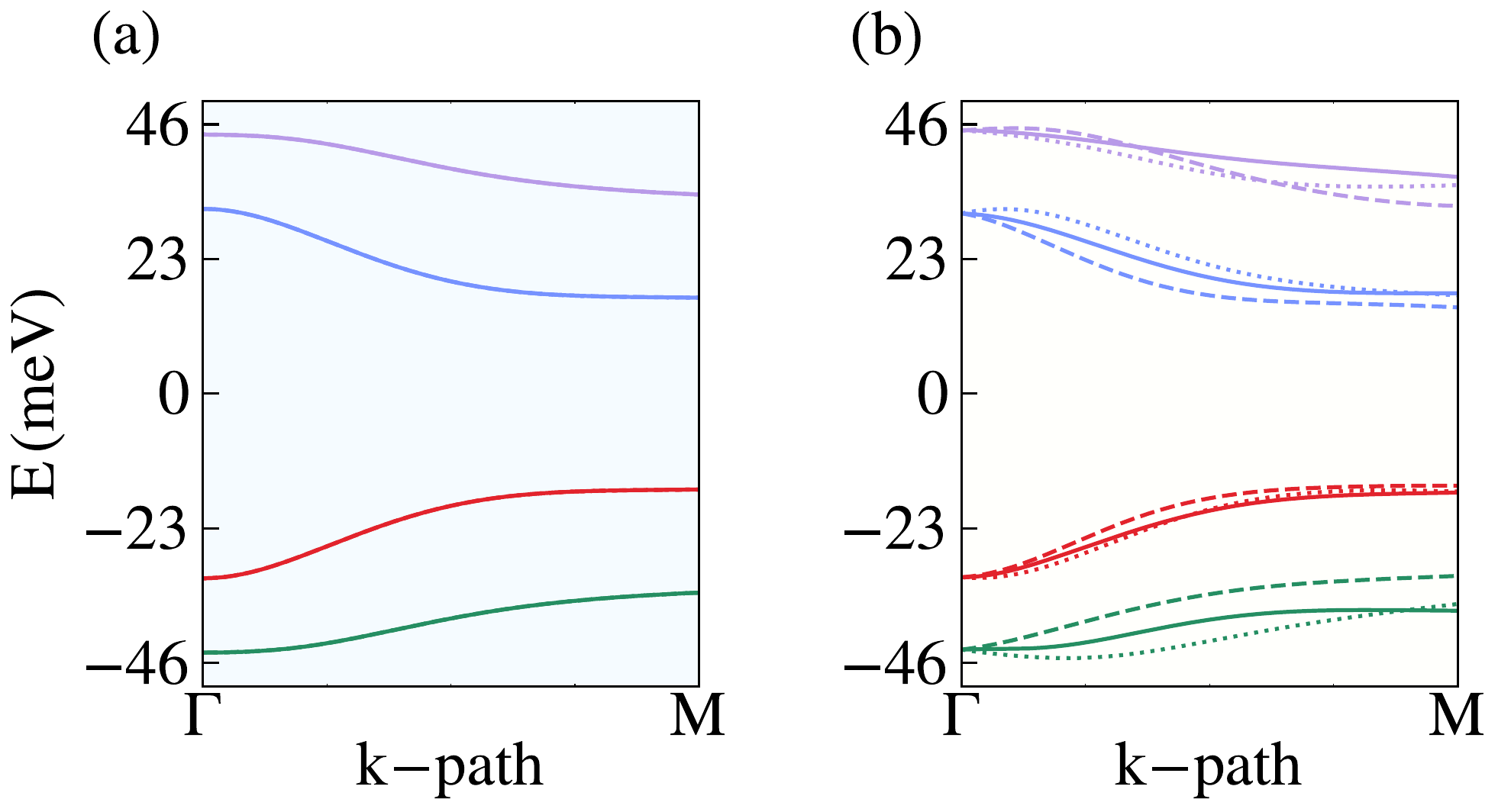}
    \caption{Band structures of the \textit{twisted bilayer} WM lattice at a twist angle of $\theta=1.40^\circ$ with $\Delta=38$~meV in the chiral limit of the system. Panel (a) shows the energy spectrum in the absence of strain, whereas panel (b) displays the energy dispersion under a uniaxial strain of magnitude $\epsilon_p=0.2\%$ applied along the zigzag direction ($\beta=0$).}
    \label{fig:Figure2}
\end{figure}
\par Now to demonstrate the breaking of the $\mathcal{C}_{3z}$ symmetry by strain, we examine the electronic band structure of the \textit{twisted bilayer} WM lattice in the chiral limit (by setting $u_1 = u_3 = u_4 = 0$, $u_2 = 110.7$~meV and $u_2^{\prime} = \frac{\sqrt{3}}{2}u_2$), at a twist angle of $\theta=1.40^\circ$, which is close to the magic angle and ensures that the four band subspace is well separated from the rest of the bulk bands.
Fig.~\ref{fig:Figure2}(a) and Fig.~\ref{fig:Figure2}(b) displays the energy dispersion in the absence of strain and in the presence of a uniaxial strain of magnitude $\epsilon_p = 0.2\%$ applied along the zigzag direction ($\beta=0$), respectively.
We compute the energy dispersions along the three equivalent high symmetry paths, $\Gamma$–M$_1$, $\Gamma$–M$_2$, and $\Gamma$–M$_3$, of the unstrained MBZ. The corresponding band dispersions along these equivalent paths are represented by solid, dashed, and dotted curves, respectively. As shown in Fig.~\ref{fig:Figure2}(a), the band structures along the three $\mathcal{C}_{3z}$ related high symmetry paths perfectly lie on top of each other in the absence of strain, indicating that the $\mathcal{C}_{3z}$ symmetry of the system is preserved. In contrast, Fig.~\ref{fig:Figure2}(b) shows that these bands exhibit significant deviations from one another along the same $\mathcal{C}_{3z}$ symmetric paths under the application of strain, thereby confirming the breaking of the $\mathcal{C}_{3z}$ symmetry in the \textit{twisted bilayer} WM lattice.
\begin{figure}[h]
    \centering
    \includegraphics[width=1.0\linewidth]{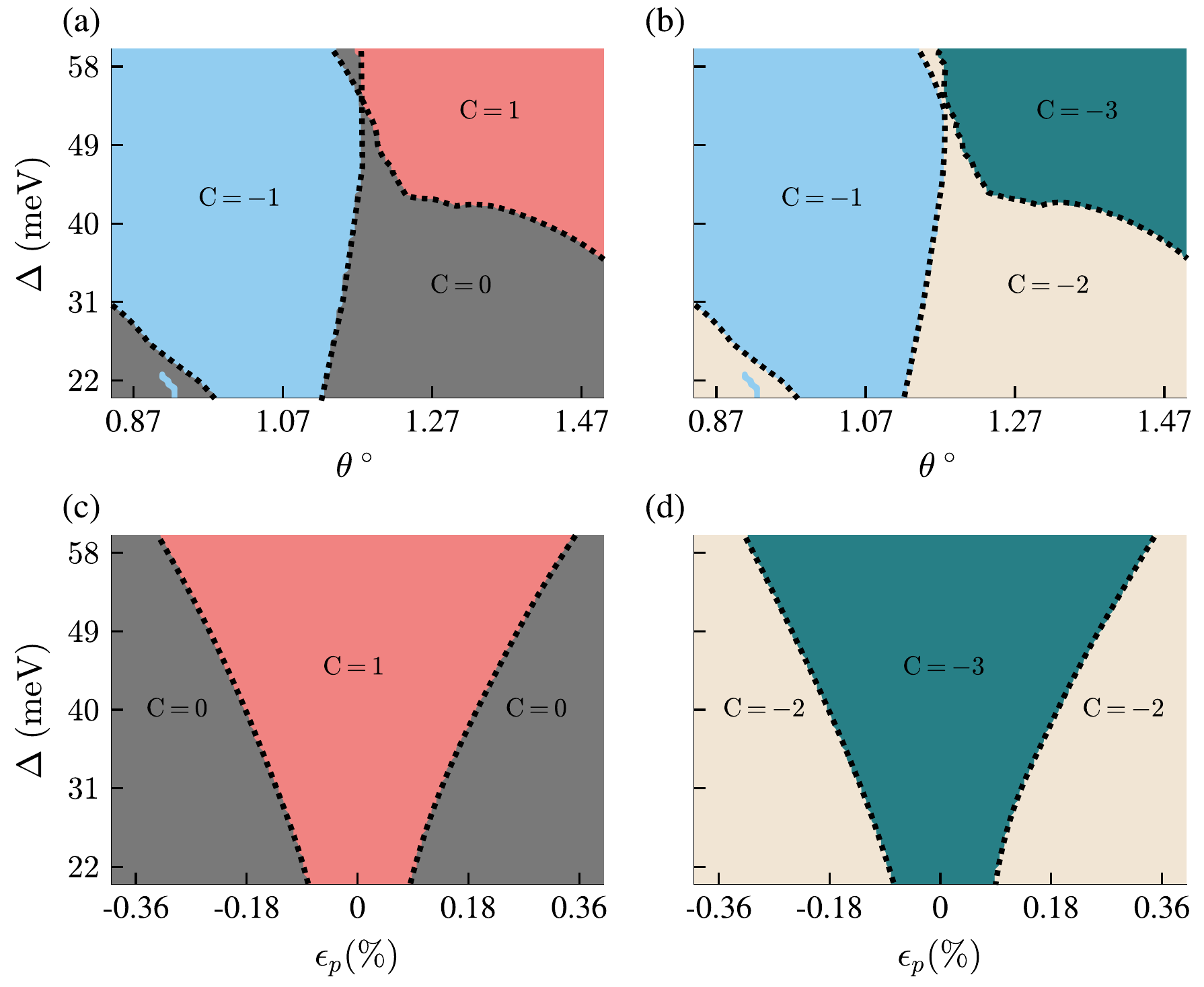}
    \caption{Panels (a) and (b) show the Chern phase diagrams of the two valence bands with band indices $n=322$ and $323$, respectively, closest to the charge neutrality point in the $\Delta$-$\theta$ plane for a fixed strain of $\epsilon_p=0.2\%$ within the chiral limit of the system. Panels (c) and (d) present the corresponding Chern phase diagrams of these bands in the $\Delta$-$\epsilon_p$ plane at a fixed twist angle of $\theta=1.40^\circ$.}
    \label{fig:Figure11}
\end{figure}
\begin{figure*}
    \centering
    \includegraphics[width=1\textwidth]{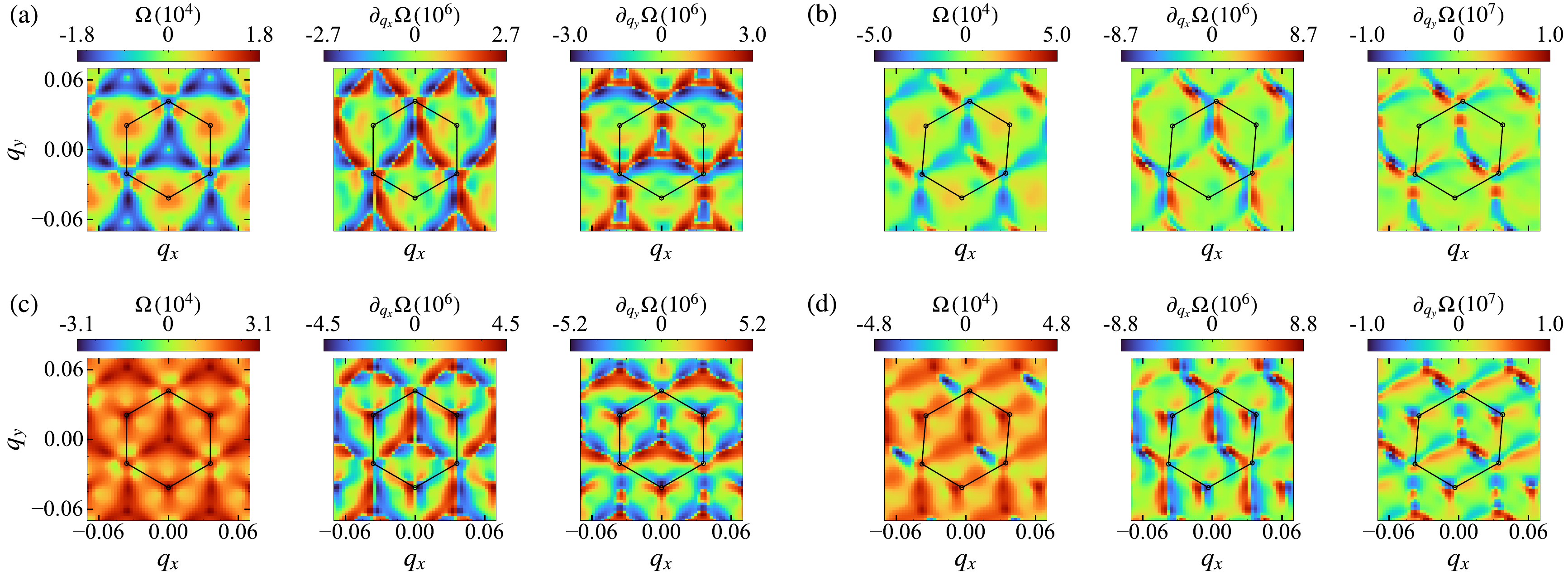}
    \caption{Panels (a) and (c) represent the distributions of the Berry curvature ($\Omega_z$) and the BCD density ($\partial_{q_b}\Omega_z$ with $b=x,y$) around the $K_+$ valley in the absence of strain for the two valance bands closest to charge neutrality, corresponding to band indices $n=322$, and $323$, respectively. Panels (b) and (d) show the corresponding distributions after applying a uniaxial strain of magnitude $\epsilon_p=0.2\%$. All results are obtained at a twist angle of $\theta=1.40^\circ$ and an onsite potential of $\Delta=38$~meV, within the chiral limit by setting $u_1=u_3=u_4=0$, $u_2=110.7$~meV, and $u_2^{\prime}=\frac{\sqrt{3}}{2}u_2$.}
    \label{Fig:Figure3}
\end{figure*}
\par Having established the strain induced breaking of the $\mathcal{C}_{3z}$ symmetry, we now investigate the topological phases associated with the $K_+$ valley in the strained \textit{twisted bilayer} WM lattice. We first construct the phase diagram in the $\Delta$-$\theta$ plane and then examine its evolution in the $\Delta$-$\epsilon_p$ plane. It is important to note that our analysis is restricted to the weak strain regime ($|\epsilon_p|\leq 0.4\%$), where the first-order approximation in the strain tensor remains valid. Fig.~\ref{fig:Figure11}(a) and Fig.~\ref{fig:Figure11}(b) show the Chern phase diagrams of the two valence bands (band indices $n=322$ and $323$) closest to the Fermi energy in the $\Delta$-$\theta$ plane for a fixed strain of $\epsilon_p=0.2\%$ within the chiral limit of the system. The corresponding Chern phase diagrams of these bands in the $\Delta$-$\epsilon_p$ plane, obtained at a fixed twist angle of $\theta=1.40^\circ$, are shown in Fig.~\ref{fig:Figure11}(c) and Fig.~\ref{fig:Figure11}(d), respectively. The phase diagrams reveal several topological phases characterized by the Chern numbers $C=0$, $\pm 1$, $-2$, and $-3$, separated by distinct transition lines. Also similar to the unstrained case, these topological phase transitions are accompanied by the closing and reopening of the band gap at specific $\mathbf{q}$ points in the MBZ. We now study the emergence of BCD in the strained \textit{twisted bilayer} WM lattice and examine its evolution across different topological phases discussed above.
\section{NLHE in twisted bilayer WM lattice}
The application of an in plane ac electric field in NLHE, $\mathbf{E}(t)=\mathrm{Re}\left(Ee^{i\omega t}\right)$ with frequency $\omega$, generates a transverse Hall current, $J(t)=\mathrm{Re}\left(j^{0}+j^{2\omega}e^{i2\omega t}\right)$, consisting of a rectified (dc) component, $j^{0}_{a}=\chi_{abc}E_{b}E^{*}_{c}$, and a second harmonic component, $j^{2\omega}_{a}=\chi_{abc}E_{b}E_{c}$, oscillating at twice of the driving frequency, $2\omega$. The corresponding nonlinear Hall susceptibility is given by,
\begin{eqnarray}
\chi_{abc} = - \frac{e^3 \tau}{2 (1 + i \omega \tau)}\, \epsilon_{adc}\, \mathcal{D}_{bd},
\end{eqnarray}
where $e$ is the electronic charge, $\tau$ denotes the scattering time, $\epsilon_{adc}$ is the Levi-Civita tensor, and $\mathcal{D}_{bd}$ represents the BCD. In 2D, the BCD can be expressed as~\cite{Sodemann2015},
\begin{eqnarray}
\mathcal{D}_{bd} = \sum_{n, \zeta} \int_{\mathbf{q}} \frac{d^2q}{(2\pi)^2}f_0\, \partial_b \Omega_d,
\end{eqnarray}
where the summation is carried out over the band index $n$ and the valley index $\zeta$. Here, $\partial_b \equiv \frac{\partial}{\partial q_b}$ denotes the partial derivative with respect to the momentum component $q_b$, $f_0$ is the equilibrium Fermi-Dirac distribution function, and $\Omega_d$ represents the $z$-component of the Berry curvature ($d=z$), whose explicit form is given by~\cite{Xiao2010,Debnath2025,Debnath2025B},
\begin{equation}
\Omega_{n,z}(\mathbf{q})
=
-2\,\mathrm{Im}
\left\langle
\partial_{q_x}\psi_n(\mathbf{q})
\middle|
\partial_{q_y}\psi_n(\mathbf{q})
\right\rangle.
\end{equation}
Here, $|\psi_n(\mathbf{q})\rangle$ denotes the Bloch wave function of the $n$-th band at momentum $\mathbf{q}$. In the following analysis, we use the notation $\mathcal{D}_b \equiv \mathcal{D}_{bz}$ ($b=x,y$) to denote the in-plane components of the BCD in 2D.
\par Having introduced the BCD in the \textit{twisted bilayer} WM lattice, we now investigate the profiles of the Berry curvature, $\Omega_z(\mathbf{q})$, and the BCD density, $\partial_{q_b}\Omega_z$ ($b=x,y$), around the $K_+$ valley for both the unstrained ($\mathcal{C}_{3z}$ protected) and strained ($\mathcal{C}_{3z}$ broken) systems. Our analysis is carried out for the two valence bands closest to the Fermi energy (corresponding band indices are $n=322$ and $323$) in the chiral symmetry regime of the system. It is important to note that, as a consequence of TRS, the Berry curvature in the \textit{twisted bilayer} WM lattice obeys:
$\Omega_{z,K_+}(\mathbf{q}) = -\Omega_{z,K_-}(-\mathbf{q})$. In the presence of $\mathcal{C}_{3z}$ symmetry, i.e., in the absence of strain, both the Berry curvature and the BCD density exhibit symmetric distributions throughout the MBZ, as illustrated in Fig.~\ref{Fig:Figure3}(a) and Fig.~\ref{Fig:Figure3}(c). Interestingly, although each valley of the \textit{twisted bilayer} WM lattice hosts a substantial Berry curvature, we find that the total BCD obtained by integrating the BCD density over the entire MBZ, vanishes identically in the absence of strain. However, in realistic materials strain breaks the $\mathcal{C}_{3z}$ symmetry. As a result, the BCD density no longer retains its symmetric profile and instead becomes asymmetric, as evident from Fig.~\ref{Fig:Figure3}(b) and Fig.~\ref{Fig:Figure3}(d), resulting in a finite dipole within the moiré unit cell.
\begin{figure}[h]
    \centering
    \includegraphics[width=0.48\textwidth]{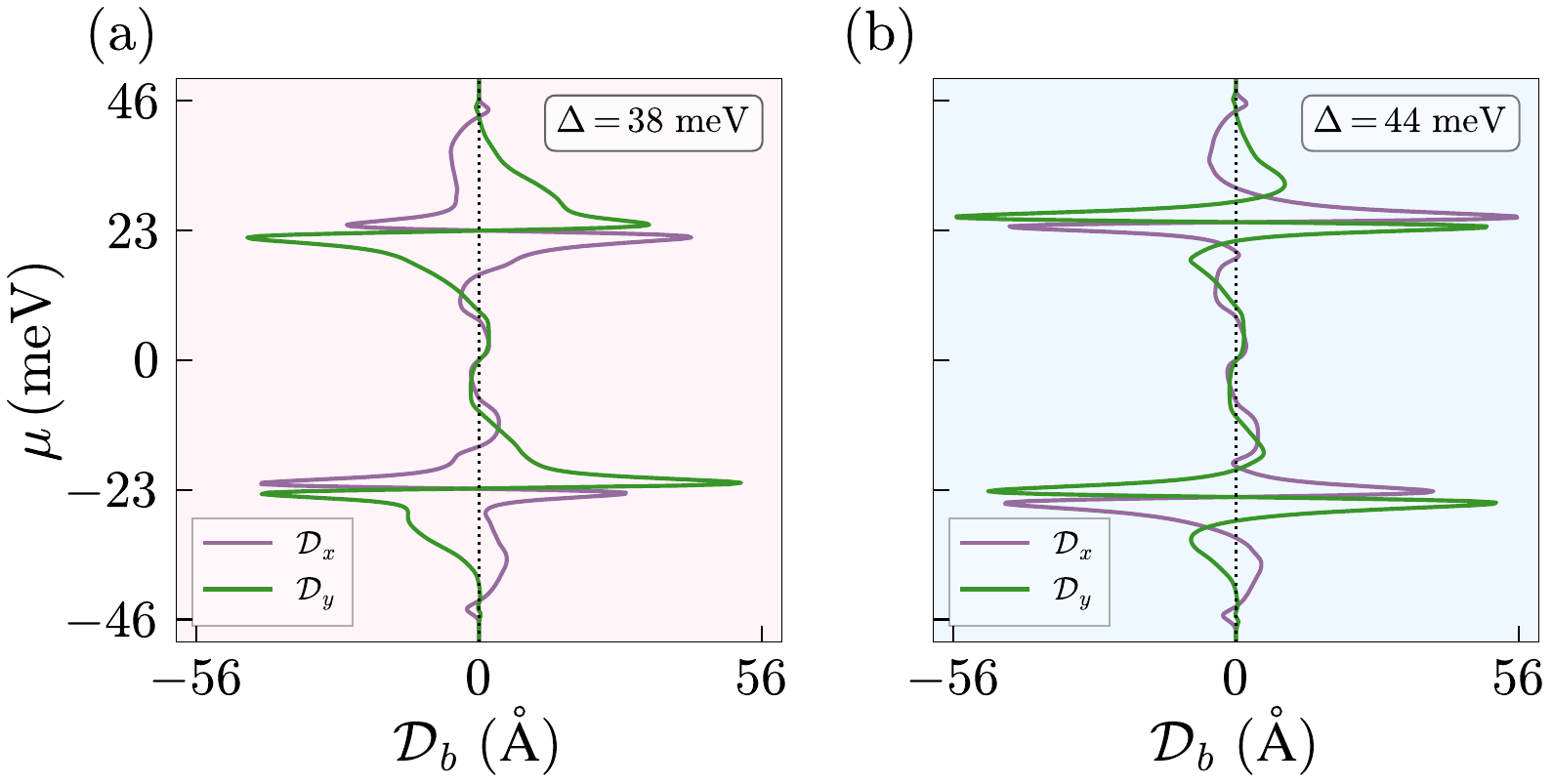}
    \caption{Sign reversal of the BCD components, $\mathcal{D}_x$ and $\mathcal{D}_y$, at a fixed strain of $\epsilon_p=0.2\%$, twist angle $\theta=1.40^\circ$, and temperature $T=4$~K. Panels (a) and (b) correspond to onsite potentials of $\Delta=38$~meV and $\Delta=44$~meV, respectively, associated with the Chern phase diagrams of the valence bands shown in Figs.~\ref{fig:Figure11}(a) and \ref{fig:Figure11}(b). The sign reversals occur in the negative (positive) chemical potential regime, indicating that the associated band inversion takes place between the two valence (conduction) bands.}
    \label{Fig:Figure4}
\end{figure}
\section{Probing topological phase transitions} 
Having discussed the Berry curvature and BCD density of the two valence bands closest to the Fermi energy in both the unstrained and strained regimes, we now investigate the evolution of the BCD across the topological phase transitions presented in Fig.~\ref{fig:Figure11}[(a)-(d)]. Our analysis is restricted to the range of chemical potential over which the BCD components associated with these bands exhibit most pronounced sign reversals. Further, the in-plane components of BCD, $\mathcal{D}_x$ and $\mathcal{D}_y$, are multiplied by an overall factor of $4$ to account for the spin and valley degeneracies. We now examine these dipole components at a fixed strain of $\epsilon_p=0.2\%$ and a twist angle $\theta=1.40^\circ$, considering two onsite potentials, $\Delta=38$~meV and $\Delta=44$~meV, corresponding to the Chern phase diagrams shown in Fig.~\ref{fig:Figure11}(a) and Fig.~\ref{fig:Figure11}(b). It is observed that both dipole components, $\mathcal{D}_x$ and $\mathcal{D}_y$, undergo clear sign reversals across the associated topological phase transitions. Notably, these reversals occur in the negative (positive) chemical potential regime, indicating that the associated band inversion takes place between the two valence (conduction) bands. The experimentally accessible sign reversals of the BCD serve as clear signatures of band inversion, enabling the identification of topological phase transitions in time reversal symmetric materials with broken inversion and $\mathcal{C}_{3z}$ symmetries. It is important to note that applying a uniaxial strain along the zigzag direction ($\beta=0$) of the bottom layer, which is subsequently twisted, induces a BCD component perpendicular to the strain direction. Consequently, both the in-plane components, namely $\mathcal{D}_x$ and $\mathcal{D}_y$, yield finite values.
\par
\begin{figure}[h]
    \centering
    \includegraphics[width=0.48\textwidth]{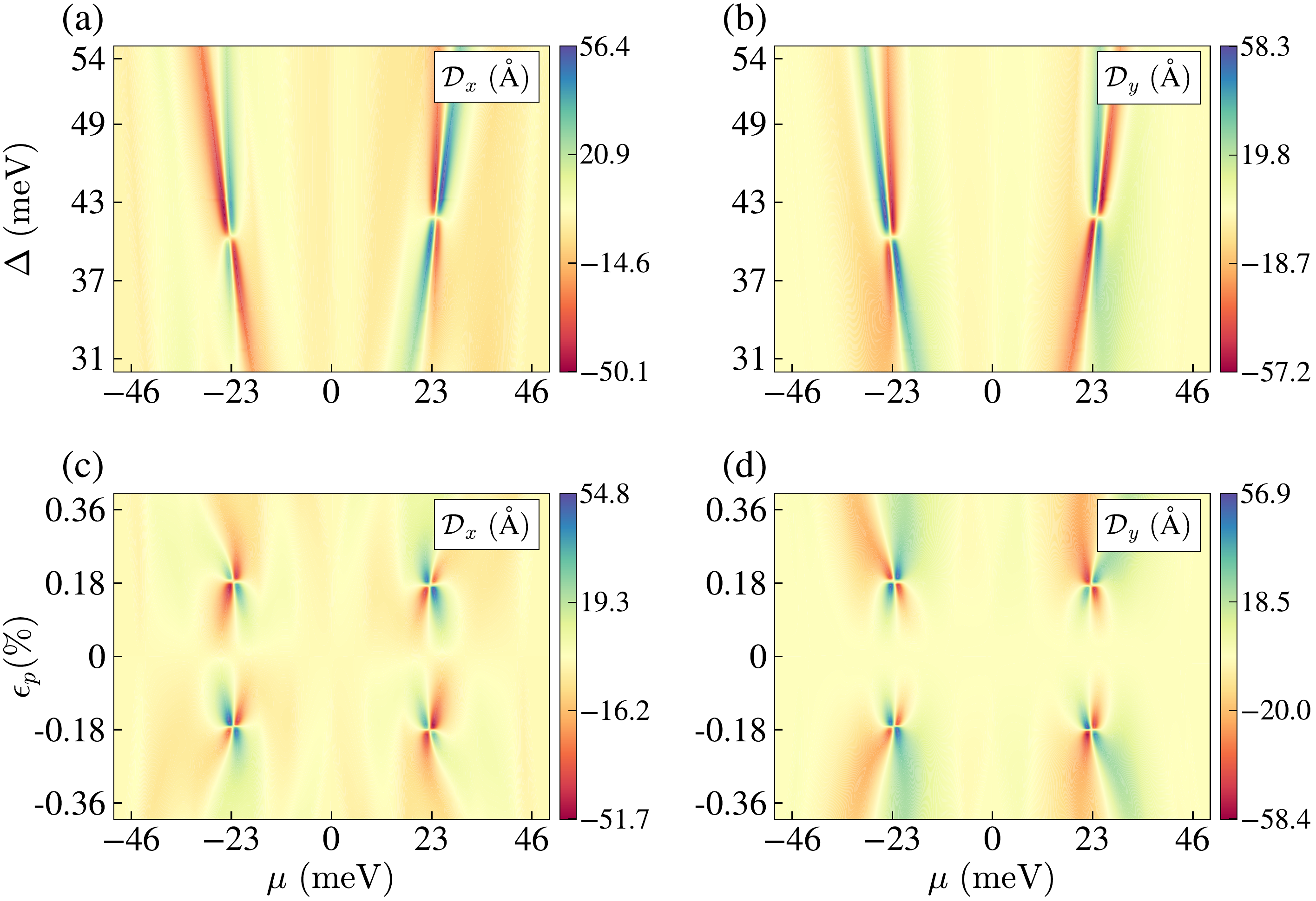}
    \caption{Topological phase transitions identified through the evolution of the BCD components, $\mathcal{D}_x$ and $\mathcal{D}_y$. Panels (a) and (b) show the corresponding phase maps in the $\Delta$-$\mu$ plane at fixed twist angle of $\theta=1.40^\circ$ and a strain strength $\epsilon_p=0.2\%$. Panels (c) and (d) present the BCD response in the $\epsilon_p$-$\mu$ plane for a twist angle of $\theta=1.40^\circ$, an onsite potential $\Delta=38$~meV and temperature $T=4$~K.}
    \label{Fig:Figure5}
\end{figure}

 Now to further elucidate the sign reversal of the BCD components across the topological phase transitions, first we examine the evolution of BCD components, $\mathcal{D}_x$ and $\mathcal{D}_y$, in the $\Delta$-$\mu$ parameter space at a fixed twist angle of $\theta=1.40^\circ$ and an applied strain $\epsilon_p=0.2\%$, as presented in Fig.~\ref{Fig:Figure5}(a) and Fig.~\ref{Fig:Figure5}(b), respectively. The butterfly like features observed in the colormaps of $\mathcal{D}_x$ and $\mathcal{D}_y$ mark the locations of the topological phase transitions and can be directly correlated with the Chern phase diagrams shown in Fig.~\ref{fig:Figure11}(a) and Fig.~\ref{fig:Figure11}(b). For a fixed value of $\Delta$, a horizontal cut through the lower lobes of each butterfly structure captures the variation of $\mathcal{D}_x$ and $\mathcal{D}_y$ as a function of the chemical potential $\mu$ for the corresponding band. In contrast, a vertical cut at a fixed $\mu$ reveals the sign reversal of the BCD components as a function of $\Delta$, thereby tracking the evolution of the same band across the phase transition. The colormaps in Fig.~\ref{Fig:Figure5}[(a)-(b)] reveal that topological phase transition occurs between the two valence (conduction) bands. In particular, the Chern number of band $n=322$ changes from $C=0$ $\rightarrow$ $C=1$, while that of band $n=323$ evolves from $C=-2$ $\rightarrow$ $C=-3$. A direct comparison between Fig.~\ref{fig:Figure11}[(a)–(b)] and Fig.~\ref{Fig:Figure5}[(a)–(b)] shows that the transition between two valence (conduction) bands occur at $\Delta=40.3$~meV ($41.8$~meV). This correspondence demonstrates that the BCD provides an experimentally relevant signature for probing the topological phase transitions.
\par Second, we investigate the BCD components, $\mathcal{D}_x$ and $\mathcal{D}_y$, in the $\epsilon_p$-$\mu$ parameter space at a fixed twist angle of $\theta=1.40^\circ$ and an onsite potential $\Delta=38$~meV, as shown in Fig.~\ref{Fig:Figure5}(c) and Fig.~\ref{Fig:Figure5}(d), respectively. For a fixed chemical potential $\mu$, a vertical cut reveals the sign reversal of $\mathcal{D}_x$ and $\mathcal{D}_y$ as functions of the strain strength $\epsilon_p$, thereby tracking the evolution of the same band across the corresponding topological phase transition.
The two butterfly like structure appearing in the chemical potential range, $-50\leq\mu\leq0$ ($0\leq\mu\leq50$), indicates that each band undergoes two successive topological phase transitions and the phase transitions occur between two valence (conduction) bands. These characteristic structures in the colormaps of $\mathcal{D}_x$ and $\mathcal{D}_y$ mark the transition points and can be directly correlated with the Chern phase diagrams shown in Fig.~\ref{fig:Figure11}(c) and Fig.~\ref{fig:Figure11}(d). Specifically, band $n=322$ undergoes the sequence of topological phase transitions as, $C=0$ $\rightarrow$ $C=1$ $\rightarrow$ $C=0$, whereas band $n=323$ evolves as, $C=-2$ $\rightarrow$ $C=-3$ $\rightarrow$ $C=-2$.
A direct comparison between Fig.~\ref{fig:Figure11}[(c)-(d)] and Fig.~\ref{Fig:Figure5}[(c)-(d)] shows that the topological phase transitions, accompanied by band-gap closings between two valence (conduction) bands, occur at $\epsilon_p = -0.173\%$ and $0.183\%$ ($\epsilon_p = -0.183\%$ and $0.173\%$).
\section{Strain and temperature dependence of BCD}
 \begin{figure}[h]
    \centering
    \includegraphics[width=0.475\textwidth]{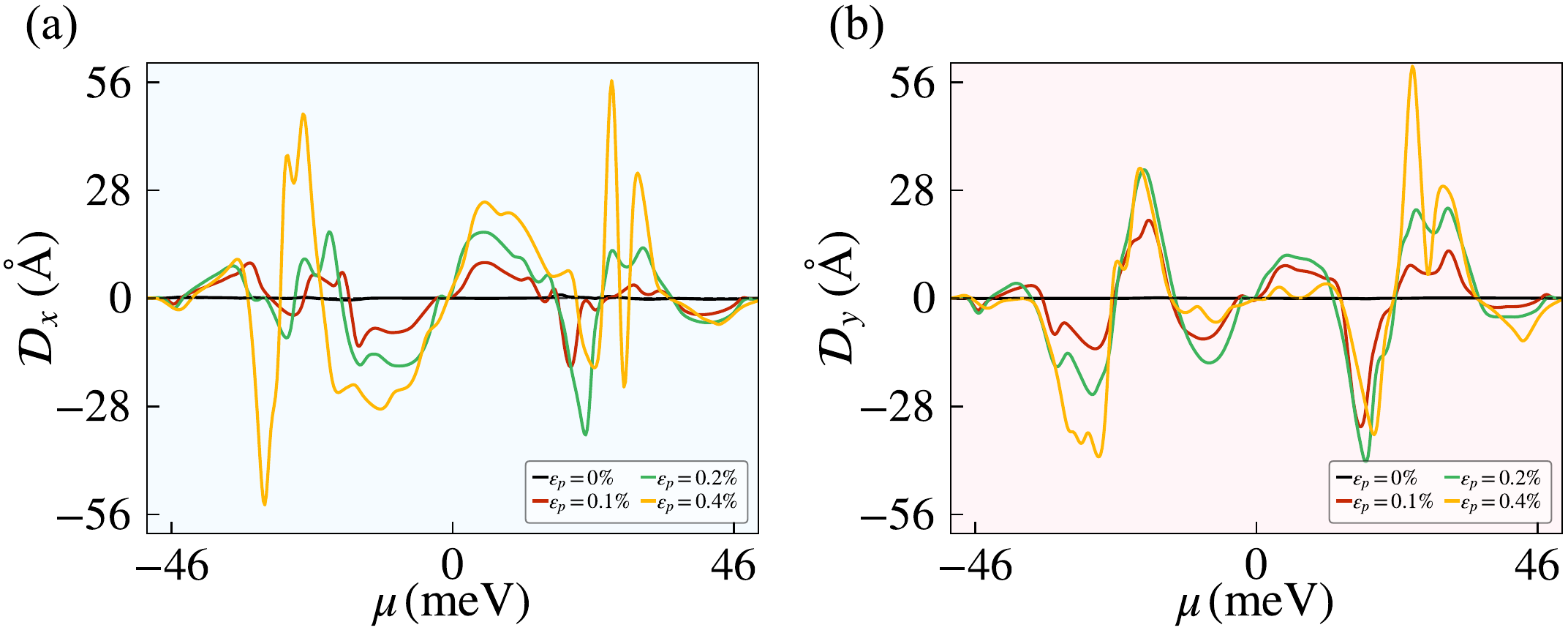}
    \caption{Strain dependency of the BCD components, $\mathcal{D}_x$ and $\mathcal{D}_y$, are presented in panels (a) and (b), respectively. The dipole response is evaluated for five different strain values, $\epsilon_p=0\%$, $0.1\%$, $0.2\%$, and $0.4\%$. BCD increases with increasing strain strength, whereas both the dipole components vanish in the absence of strain due to the preserved $\mathcal{C}_{3z}$ symmetry of the system.}
    \label{Fig:Figure6}
\end{figure}
We now examine the gate-dependent behavior of the BCD components, $\mathcal{D}_x$ and $\mathcal{D}_y$, at a fixed twist angle $\theta=1.40^\circ$, an onsite potential $\Delta=38$~meV, and temperature $T=4$~K, under different values of the uniaxial strain applied along the zigzag direction ($\beta=0$) in the bottom layer of the \textit{twisted bilayer} WM lattice. The results are obtained in the broken chiral symmetry regime by considering the interlayer coupling parameters as, $u_2=110.7$~meV, $u_1=u_3=u_4=0.2~u_2$, and $u_2^{\prime}=\frac{\sqrt{3}}{2}u_2$. The resulting variations of $\mathcal{D}_x$ and $\mathcal{D}_y$ are presented in Figs.~\ref{Fig:Figure6}(a) and \ref{Fig:Figure6}(b), respectively. It is evident from Fig.~\ref{Fig:Figure6} that the BCD components, $\mathcal{D}_x$ and $\mathcal{D}_y$, exhibit pronounced oscillations near the Fermi energy and undergo multiple sign reversals within a narrow window of the chemical potential $\mu$. The maximum value of the BCD emerges in the vicinity of the band anticrossing points. Notably, the maximum value of the BCD reaches approximately $58$~\AA\ in the chiral limit [see Fig.~\ref{Fig:Figure5}(d)] of the system and increases to about $61$~\AA\ in the weakly broken chiral symmetry regime [see Fig.~\ref{Fig:Figure6}(b)]. Moreover, the strain dependence reveals a gradual enhancement of the dipole strength with increasing of strain values, whereas the BCD components vanish throughout the entire chemical potential range ($-50<\mu<50$) in the absence of strain.
This further highlights the crucial role of breaking the $\mathcal{C}_{3z}$ symmetry in generating a finite BCD in the \textit{twsited bilayer} WM lattice. Furthermore, the dip-to-peak and the peak-to-dip evolution of $\mathcal{D}_x$ and $\mathcal{D}_y$ with chemical potential $\mu$ can be attributed to the change in sign of the Berry curvature density in the vicinity of the band anticrossing points. Consequently, the BCD components undergo sign reversals as the chemical potential $\mu$ is swept across the relevant bands. We re-emphasize here that our analysis is restricted to the bands near the Fermi energy of the system, as the degeneracy of the bands away from the charge neutrality precludes a meaningful analysis of the BCD response. 
\begin{figure}[h]
    \centering
    \includegraphics[width=0.482\textwidth]{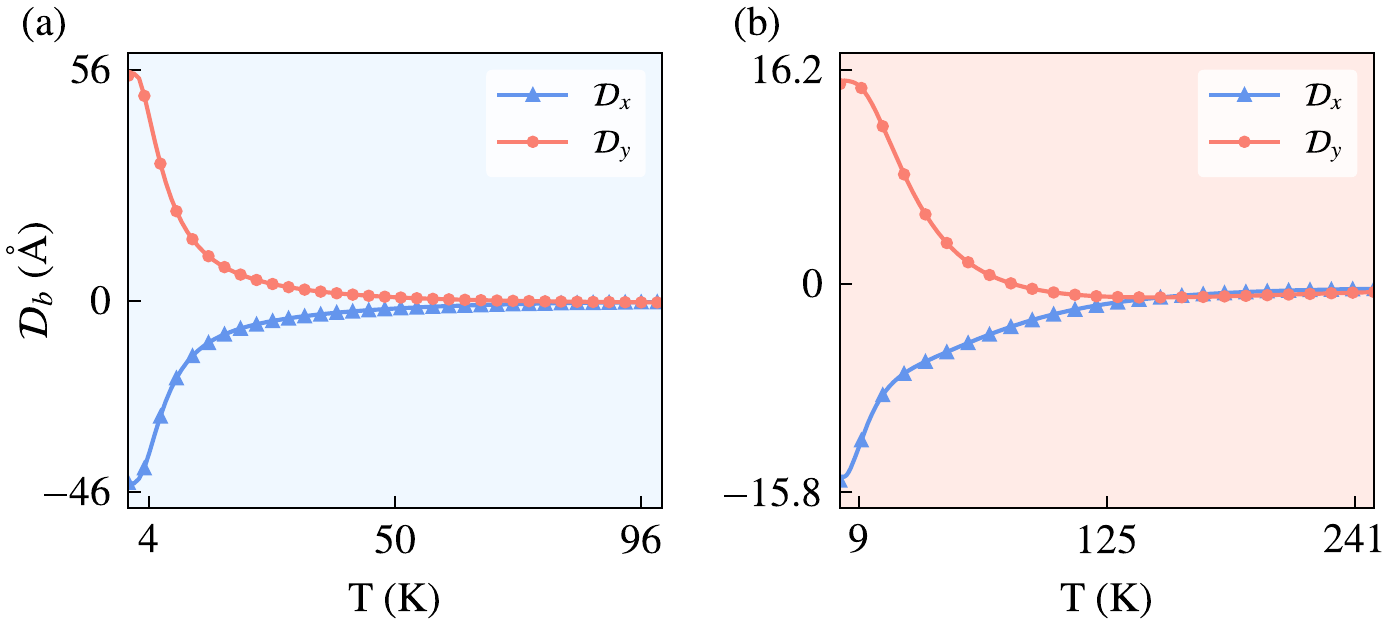}
    \caption{Temperature dependence of the BCD components at fixed values of the twist angle, strain strength, and onsite potential: $\theta=1.08^\circ$, $\mathcal{E}_p=0.2\%$, and $\Delta=38$~meV, respectively. Panel (a) presents the results in the chiral limit, whereas panel (b) corresponds to the broken chiral symmetry regime. The chemical potential is set to $\mu=-21.7$~meV in panel (a) and $\mu=-15.7$~meV in panel (b).}
    \label{Fig:Figure7}
\end{figure}
\par So far, all the results discussed above have been obtained at a fixed temperature $T=4$~K. We now investigate the temperature dependence of the BCD in the \textit{twisted bilayer} WM lattice by examining the variation of the dipole components, $\mathcal{D}_x$ and $\mathcal{D}_y$, with temperature. The results are presented in Fig.~\ref{Fig:Figure7}(a) for the chiral limit and in Fig.~\ref{Fig:Figure7}(b) for the broken chiral symmetry regime. Throughout this analysis, the twist angle, strain strength, and onsite potential are fixed at $\theta=1.40^\circ$, $\epsilon_p=0.2\%$, and $\Delta=38$~meV, respectively.
It is clearly seen from Fig.~\ref{Fig:Figure7} that the magnitudes of the BCD components gradually decrease with increasing temperature in both the chiral and broken chiral symmetry regimes. This behavior can be attributed to the thermal broadening of the Fermi-Dirac distribution, which albeit being small, has profound effect in diminishing the slope and consequently suppressing the BCD response.
This behavior suggests that the \textit{twisted bilayer} WM lattice can exhibit a large NLH response at low temperatures. Moreover, the temperature dependence of the BCD components follows trends similar to those reported in the earlier studies of TBG~\cite{Duan2022}, TBDL~\cite{paul2026arXiv}, and twisted bilayer WTe$_2$~\cite{Cao2025}.
\section*{Summary and conclusion}
We theoretically investigate the topological phase transitions corresponding to the bands near the charge neutrality in the \textit{twisted bilayer} WM lattice under the effect of an onsite mass term that breaks the inversion symmetry. To begin with, we analyze the unstrained system which exhibits certain sectors of high Chern numbers that can be related to the elevated Wilson loop winding of the isolated flat bands of the WM lattice. Subsequently, we investigate how strain induced breaking of the $\mathcal{C}_{3z}$ symmetry modifies the corresponding topological phases. The simultaneous breaking of inversion and $\mathcal{C}_{3z}$ symmetries gives rise to a finite BCD, which provides an experimentally accessible probe for identifying topological phase transitions in time reversal symmetric systems, where the underlying band geometry is otherwise difficult to characterize. We show that the BCD components correctly track the topological phase transitions of the relevant bands through a complete reversal of their signs across the transition boundary. The associated sign change in both the Berry curvature and the dipole provides a clear signature of band inversion driven by the closing and reopening of the band gap. We also analyze the evolution of the BCD strength with the applied strain in the broken chiral symmetry regime. The BCD remains nearly negligible at weak strain values and gradually increases with increasing strain strength, highlighting the essential role of $\mathcal{C}_{3z}$ symmetry breaking in generating a finite BCD. The calculated BCD magnitude in \textit{twisted bilayer} WM lattice reaches approximately $58$~\AA\ in the chiral limit and is further enhanced to about $61$~\AA\ in the broken chiral symmetry regime. 
\par For comparison, in other strained materials, such as strained TBG~\cite{Battilomo2019} and strained MoS$_2$~\cite{Son2019}, the strain-induced BCD is typically of the order of $\sim 10^{-2}$~\AA. In contrast, bilayer and multilayer WTe$_2$ systems~\cite{Du2018, Wang2019} exhibit larger BCD values, with magnitudes of approximately $0.2$~\AA\ and $1$~\AA, respectively. In twisted bilayer systems, the BCD can assume substantially larger values. Specifically, TBG~\cite{Zhang2022}, twisted bilayer WSe$_2$~\cite{Hu2022}, and spin-orbit coupled twisted bilayer WTe$_2$~\cite{He2021} exhibit dipole magnitudes of approximately $200$~\AA, $3$~\AA, and $1500$~\AA, respectively. We additionally study the BCD components in both the chiral limit and broken chiral symmetry regimes. As the temperature increases, the BCD components gradually decrease. This behavior indicates that one could detect a strong NLH response at very low temperatures in \textit{twisted bilayer} WM lattice. 
\par Furthermore, we want to point out that our results are well converged with respect to the number of lattice points in the reciprocal space to ensure the authenticity of our findings. It is important to emphasize that the contributions to the NLHE can be decomposed into intrinsic (geometric) and extrinsic (disorder included) contributions~\cite{Du2021}. In our work, we restrict our analysis to the intrinsic contribution of the NLH response, which is directly related to the BCD. 
It is important to mention here that, as long as the $\mathcal{C}_{3z}$, $\mathcal{C}_{2z}\mathcal{T}$, and chiral symmetries are preserved in the \textit{twisted bilayer} WM lattice, small variations in the hopping amplitudes do not lift the fourfold degeneracies at the valleys. In particular, the hopping amplitude between the $B$ and $C$ orbitals, $t$, is $2/\sqrt{3}$ times larger than those involving the other orbital pairs, denoted by $t_{\text{other}}$. However, varying the ratio $t/t_\text{other}$ does not lift the degeneracy; instead, it modifies the velocity of the Dirac cones while preserving the nontrivial winding of the pseudospin texture around each valley. This robustness against variations in the hopping amplitudes provides a favorable route toward realizing materials with low-energy states characterized by a pseudospin-$3/2$ structure. A promising candidate for realizing the watermill lattice is the LaAlO$_3$/SrTiO$_3$ (111) quantum well~\cite{Doennig2013}. Other viable platforms for engineering this lattice include covalent-organic and metal-organic frameworks~\cite{Ni2022,Niu2022,Pan2023,Yan2021,Ni2022C,Zhang2023B,Hu2022B,Jiang2021}, as well as CO molecules adsorbed on a Cu(111) surface~\cite{Tassi2024,Gomes2012,Khajetoorians2019}. Other potential material platforms include two-dimensional group-IV systems, such as germanene and stanene \cite{Barman2024, Zhang2023, Acun2015}, as well as MXenes \cite{Li2023}, where the relevant orbitals may not lie in the same plane, but the hopping configuration essentially resembles what has been discussed in our work.
\section*{Acknowledgments}
SL and GP sincerely thank Ms. Shreya Debnath for her valuable comments and suggestions. SL acknowledges financial support from the MoE, Govt. of India, through the Prime Minister’s Research Fellowship (PMRF) scheme in May 2022. SL and GP further acknowledge the National Supercomputing Mission (NSM) for providing computing resources of `PARAM Kamrupa' at IIT Guwahati, which is implemented by C-DAC and supported by the Ministry of Electronics and Information Technology (MeitY) and Department of Science and Technology (DST), Government of India. BT is supported by The Scientific and Technological Research Council of T{\"u}rkiye (T{\"U}B{\.I}TAK) under Grant No. 125F435 and Turkish Academy of Sciences (TUBA) under Grant No. AD-2026. SB acknowledges 
support from T{\"U}B{\.I}TAK-BIDEB.

\section*{Data availability} 
The data that support the findings of this work are not publicly available. The data are available from the authors upon reasonable request.

\appendix
\section{TIGHT BINDING MODEL OF MONOLAYER WM LATTICE}\label{Appendix-A}
The Bravais lattice vectors of the WM lattice are given by $\mathbf{a}_1 = a \left( \frac{1}{2}, \frac{\sqrt{3}}{2} \right)$ and $\mathbf{a}_2 = a \left( -\frac{1}{2}, \frac{\sqrt{3}}{2} \right)$, where $a = \sqrt{3}d$ is the lattice constant and $d$ is the NN bond length. Similar to graphene, we take $d=1.42$~\AA. The NN hopping amplitude between the $B$ and $C$ orbitals is denoted by $t$, whereas the hopping amplitudes between the $A_1$ and $B$ orbitals and between the $A_2$ and $C$ orbitals are taken to be equal and are denoted by $t^\prime=\frac{\sqrt{3}}{2}t$.
The corresponding reciprocal lattice vectors are given by,
$\mathbf{b}_1 = \frac{4\pi}{\sqrt{3}a} \left( \frac{\sqrt{3}}{2}, \frac{1}{2} \right)$ and
$\mathbf{b}_2 = \frac{4\pi}{\sqrt{3}a} \left( -\frac{\sqrt{3}}{2}, \frac{1}{2} \right)$.
The two inequivalent valleys, $K_+$ and $K_-$, are located at $\left( \frac{4\pi}{3a}, 0 \right)$ and $\left( -\frac{4\pi}{3a}, 0 \right)$, respectively. In the presence of NN hoppings, the momentum-space Hamiltonian in the sublattice basis $\Psi$ = $\left( A_{1}^{\mathbf{k}} \quad B^{\mathbf{k}} \quad C^{\mathbf{k}} \quad A_{2}^{\mathbf{k}}\right)^{T}$ can be written as~\cite{Hung2025},
\begin{equation}
    H_{\text{WM}}(\mathbf{k}) =
\begin{pmatrix}
0 & \frac{\sqrt{3}}{2} g(\mathbf{k}) & 0 & 0 \\
\frac{\sqrt{3}}{2} g^*(\mathbf{k}) & 0 & g(\mathbf{k}) & 0 \\
0 & g^*(\mathbf{k}) & 0 & \frac{\sqrt{3}}{2} g(\mathbf{k})\\
0 & 0 & \frac{\sqrt{3}}{2} g^*(\mathbf{k}) & 0
\end{pmatrix} \label{WM_Ham},
\end{equation}
where the off-diagonal term is defined as, $g(\mathbf{k}) = - t(1 + e^{-i \mathbf{k}.\mathbf{a}_1} + e^{-i\mathbf{k}.\mathbf{a}_2})$.
\par The low-energy electronic states of the WM lattice are characterized by fourfold-degenerate Dirac cones located at the two valleys, $K_+$ and $K_-$. Near the $K_\zeta$ valley, the low-energy expansion of $g(\textbf{k})$ yields,
\begin{equation}
 g(q_x, q_y) = \frac{\sqrt{3} a t}{2} (\zeta q_x - i q_y) = v_F (\zeta q_x - i q_y), \label{g(k)}  
\end{equation}
where $\zeta = \pm$ indicates the valley index, $v_F = \frac{\sqrt{3} a t}{2}$ is the Fermi velocity and $\mathbf{q}$ is measured relative to the $K_\zeta$ point. Substituting Eq.~\eqref{g(k)} into Eq.~\eqref{WM_Ham}, we obtain the effective low-energy Hamiltonian as~\cite{Hung2025},
\begin{equation}
H^\zeta(q_x,q_y) = v_F (\zeta q_x S_1 + q_y S_2).
\end{equation}
Here, 
\begin{align}
S_1 &=
\begin{pmatrix}
0 & \dfrac{\sqrt{3}}{2} & 0 & 0\\
\dfrac{\sqrt{3}}{2} & 0 & 1 & 0\\
0 & 1 & 0 & \dfrac{\sqrt{3}}{2}\\
0 & 0 & \dfrac{\sqrt{3}}{2} & 0
\end{pmatrix},~\text{and}\nonumber\\
S_2 &=
\begin{pmatrix}
0 & -i\dfrac{\sqrt{3}}{2} & 0 & 0\\
i\dfrac{\sqrt{3}}{2} & 0 & -i & 0\\
0 & i & 0 & -i\dfrac{\sqrt{3}}{2}\\
0 & 0 & i\dfrac{\sqrt{3}}{2} & 0
\end{pmatrix}.
\label{eq:spin_matrices}
\end{align}
\section{INTRALAYER AND INTERLAYER HAMILTONIAN FOR STRAINED \textit{TWISTED BILAYER} WM LATTICE}\label{Appendix-B}
Now we construct the low-energy intralayer Hamiltonian of the constituent WM monolayers by including the NN hopping together with the effect of an onsite mass term that emulates a potential gradient (incorporated through $V_{t/b}$). The resulting Hamiltonian describes massless Dirac fermions, where the top and the bottom layers are rotated by angles $+\theta/2$ and $-\theta/2$, respectively, and given by,
\begin{eqnarray}
H_{t/b}^{\zeta}({\pm\theta/2})= v_F \mathcal{R}_{\mp \theta/2}\boldsymbol{q}\cdot(\zeta S_1,S_2) + V_{t/b},
\end{eqnarray}
where $v_F = 5960$~meV$\cdot$\AA, and $\mathcal{R}_{\theta}$ denotes the rotational operator. The momentum $\mathbf{q} = \mathbf{k} - \mathbf{K}_{\zeta}$ is defined relative to the original BZ corner $\mathbf{K}_{\zeta}$ of the rotated monolayer WM lattice and $S_i$ ($i = 1, 2, 3$) denotes the matrix representation of the spin-3/2 operators in the $S_3$ eigenbasis. The onsite potential matrix $V_{t/b}$ is given by,
\begin{eqnarray}
V_t &=& \begin{pmatrix}
\Delta_{A_1}^t & 0 & 0 & 0\\
0 & 0 & 0 & 0 \\
0 & 0 & 0 & 0 \\
0 & 0 & 0 & \Delta_{A_2}^t
\end{pmatrix} = \begin{pmatrix}
\frac{3}{2} \Delta & 0 & 0 & 0\\
0 & 0 & 0 & 0 \\
0 & 0 & 0 & 0 \\
0 & 0 & 0 & \frac{1}{2}\Delta
\end{pmatrix}, \nonumber \\
&\text{and}&\nonumber\\
V_b &=& \begin{pmatrix}
\Delta_{\tilde{A}_1}^b & 0 & 0 & 0\\
0 & 0 & 0 & 0 \\
0 & 0 & 0 & 0 \\
0 & 0 & 0 & \Delta_{\tilde{A}_2}^b
\end{pmatrix} = \begin{pmatrix}
-\frac{1}{2} \Delta & 0 & 0 & 0\\
0 & 0 & 0 & 0 \\
0 & 0 & 0 & 0 \\
0 & 0 & 0 & -\frac{3}{2}\Delta
\end{pmatrix}.
\end{eqnarray}
We incorporate the effect of strain through the linear strain tensor $\mathbf{\mathcal{E}}$, which transforms an arbitrary real space coordinate $\mathbf{r}$ according to,
\begin{equation}
\mathbf{r}^{\prime} = (\mathbb{I} + \mathbf{\mathcal{E}})
\,\mathbf{r}.
\end{equation}
The corresponding transformation in reciprocal space then follows as,
\begin{equation}
\mathbf{q}^\prime = (\mathbb{I}+ \mathbf{\mathcal{E}}^T)^{-1}\mathbf{q}
\approx
(\mathbb{I}-\mathbf{\mathcal{E}}^T)\,\mathbf{q},
\end{equation}
where, in the final expression we retain only terms up to first order in the strain tensor. Hereafter, a prime on any quantity indicates that it is evaluated in the presence of strain.
\par Now we apply a uniaxial strain of magnitude $\epsilon_p$ to the bottom layer WM lattice along a direction making an angle $\beta$ relative to the zigzag axis ($\beta=0$). The corresponding strain tensor is given by~\cite{Bi2019,Pereira2009},
\begin{eqnarray}
\mathbf{\mathcal{E}}
&=&
R_{\beta}
\begin{pmatrix}
\epsilon_p & 0 \\
0 & -\nu \epsilon_p
\end{pmatrix}
R_{\beta}^{-1}
\nonumber \\
&=&
\epsilon_p
\begin{pmatrix}
\cos^2\beta-\nu\sin^2\beta &
(1+\nu)\cos\beta\sin\beta \\
(1+\nu)\cos\beta\sin\beta &
\sin^2\beta-\nu\cos^2\beta
\end{pmatrix}.
\end{eqnarray}
Here, $\nu=0.165$ denotes the Poisson ratio of the WM lattice, for which we adopt the same value as that of graphene~\cite{Pereira2009}. The applied strain modifies both the bottom layer Hamiltonian and the interlayer tunneling as well. In the bottom layer, these strain induced effects can be incorporated through an effective gauge field~\cite{Bi2019,Sun2022}, which is given by,
\begin{equation}
\mathbf{A} = \frac{\gamma}{d}
\left( \mathcal{E}_{xx} - \mathcal{E}_{yy}\,, -2\mathcal{E}_{xy} \right),
\end{equation}
Here, $\gamma=1.57$ represents the Grüneisen parameter~\cite{Bi2019}, whose value is taken to be identical to that of graphene. Now incorporating the strain induced gauge field through the substitution $\mathbf{q}\rightarrow\mathbf{q}+\zeta\mathbf{A}$, we obtain the strained bottom layer Hamiltonian as,
\begin{eqnarray}
 H_{b}^{\prime \zeta} =  v_F \,{R}_{\theta/2} \left[ (\mathbb{I} + \mathbf{\mathcal{E}}^T)\,\mathbf{q}^{\prime} + \zeta \mathbf{A} \right].(\zeta S_1, S_2) + V_b.
\end{eqnarray}
Also the strain induced gauge field shifts the Dirac points of the bottom layer to~\cite{Bi2019},
\begin{equation}
\mathbf{\mathcal{B}}_{\zeta} = \left(\mathbb{I} - \mathbf{\mathcal{E}}^{T}\right)\,\mathbf{K}_{b,\zeta} - \zeta\,\mathbf{A}. 
\end{equation}
The applied strain also influences the interlayer tunneling by changing the momentum transfer between the two layers associated with the tunneling process. As a result, the transferred momenta become,
\begin{align}
\zeta \mathbf{q}_b^\prime &=
{R}_{-\frac{\theta}{2}} (\mathbb{I} - \mathbf{\mathcal{E}}^{T})\mathbf{K}_\zeta
- {R}_{\frac{\theta}{2}} \mathbf{K}_\zeta, \nonumber \\
\zeta\mathbf{q}_{tr}^\prime &= {R}_{-\frac{\theta}{2}} (\mathbb{I} - \mathbf{\mathcal{E}}^{T})(\mathbf{K}_\zeta + \zeta \mathbf{b}_2)
- {R}_{\frac{\theta}{2}} (\mathbf{K}_\zeta + \zeta\mathbf{b}_2), \nonumber\\
\zeta \mathbf{q}_{tl}^\prime &= {R}_{-\frac{\theta}{2}} (\mathbb{I} - \mathbf{\mathcal{E}}^{T})(\mathbf{K}_\zeta -\zeta\mathbf{b}_1)
- {R}_{\frac{\theta}{2}} (\mathbf{K}_\zeta -\zeta\mathbf{b}_1). 
\end{align}
Incorporating all these modifications into the continuum model, we obtain the strained interlayer Hamiltonian as,
\begin{eqnarray}
T^{\prime}_{\zeta}(\mathbf{q},\tilde{\mathbf{q}}) 
&=& 
T^{\prime}_{\zeta\mathbf{q}^{\prime}_b} \delta_{\mathbf{q}-\tilde{\mathbf{q}}-\zeta\mathbf{q}^{\prime}_b} + T^{\prime}_{\zeta\mathbf{q}^{\prime}_{tr}} \delta_{\mathbf{q}-\tilde{\mathbf{q}}-\zeta\mathbf{q}^{\prime}_{tr}} + T^{\prime}_{\zeta\mathbf{q}^{\prime}_{tl}} \delta_{\mathbf{q}-\tilde{\mathbf{q}}-\zeta\mathbf{q}^{\prime}_{tl}} \nonumber.\\\label{Interlayer_Coupling1}
\end{eqnarray}
Notably, the hopping matrices remain unchanged and therefore retain their original form:
\begin{eqnarray}
T^{\prime \tilde{\alpha}\delta}_{\zeta \mathbf{q}'_b} 
&=& u_{\tilde{\alpha}\delta} 
= T^{\tilde{\alpha}\delta}_{\zeta \mathbf{q}_b}, \\
T^{\prime \tilde{\alpha}\delta}_{\zeta \mathbf{q}'_{tr}} 
&=& u_{\tilde{\alpha}\delta} 
e^{i \zeta \left(-\mathbf{b}_2 \cdot \boldsymbol{\tau}_\delta 
+ \tilde{\mathbf{b}}^{\prime}_2 \cdot \boldsymbol{\tau}'_{\tilde{\alpha}} \right)} \nonumber \\
&=& u_{\tilde{\alpha}\delta} 
e^{i \zeta \left(-\mathbf{b}_2 \cdot \boldsymbol{\tau}_\delta
+ \tilde{\mathbf{b}}_2 \cdot \boldsymbol{\tau}_{\tilde{\alpha}} \right)}
= T^{\tilde{\alpha}\delta}_{\zeta \mathbf{q}_{tr}},\\
T^{\prime \tilde{\alpha}\delta}_{\zeta \mathbf{q}'_{tl}} 
&=& u_{\tilde{\alpha}\delta} 
e^{i \zeta \left(\mathbf{b}_1 \cdot \boldsymbol{\tau}_\delta 
- \tilde{\mathbf{b}}^{\prime}_1 \cdot \boldsymbol{\tau}'_{\tilde{\alpha}} \right)} \nonumber \\
&=& u_{\tilde{\alpha}\delta} 
e^{i \zeta \left(\mathbf{b}_1 \cdot \boldsymbol{\tau}_\delta 
- \tilde{\mathbf{b}}_1 \cdot \boldsymbol{\tau}_{\tilde{\alpha}} \right)}
= T^{\tilde{\alpha}\delta}_{\zeta \mathbf{q}_{tl}}.
\end{eqnarray}
To obtain these results, we used the relation $\tilde{\mathbf{b}}^{\prime}_{1,2}\cdot\boldsymbol{\tau}'_{\tilde{\alpha}}=\tilde{\mathbf{b}}_{1,2}\cdot\boldsymbol{\tau}_{\tilde{\alpha}}$, while neglecting the effect of strain on the tunneling amplitudes. Here, $\alpha,\delta \in {A_1,B,C,A_2}$ label the sublattice orbitals, with $u_{\tilde{A}_1A_1} = u_{\tilde{B}B} = u_{\tilde{C}C} = u_{\tilde{A}_2A_2} = u_1$, $u_{\tilde{B}C} = u_{\tilde{C}B} = u_2$, $u_{\tilde{A}_1B} = u_{\tilde{B}A_1} =u_{\tilde{A}_2C} = u_{\tilde{C}A_2}= u_2^\prime$, $u_{\tilde{A}_1C} = u_{\tilde{C}A_1} = u_{\tilde{A}_2B} =u_{\tilde{B}A_2}=u_3$, and $u_{\tilde{A}_1A_2} = u_{\tilde{A}_2A_1}=u_4$.
\bibliography{mainNotes}
\end{document}